\documentclass[review]{elsarticle}
\biboptions{sort&compress}

\usepackage{graphicx}
\usepackage{bm}
\usepackage{amssymb}
\usepackage{hyperref}

\begin{document}

\begin{frontmatter}

\title{Effective and efficient transport mechanism of CO$_2$
  in subnano-porous crystalline membrane of syndiotactic polystyrene}

\author{Yoshinori Tamai\corref{cor1}}
\address{Department of Applied Physics, Faculty of Engineering, 
  University of Fukui, 3-9-1 Bunkyo, Fukui 910-8507, Japan}
\ead{tamai@u-fukui.ac.jp}
\cortext[cor1]{Corresponding author}

\date{\today}

\begin{abstract}
  The gas permeation behavior in the single crystal of 
  syndiotactic polystyrene (s-PS) S-I form was investigated in detail, 
  in comparison to that in the s-PS $\varepsilon$ form.
  The S-I form exhibits high separation factors for 
  CO$_2$/N$_2$ ($\sim$180) and CO$_2$/CH$_4$ ($\sim$500), 
  while preserving its high CO$_2$ permeability 8000 Barrers;
  a trade-off between selectivity and permeability was broken through.
  The mechanism of the effective and efficient transport of CO$_2$ in
  the S-I form was examined in relation to the cavity structure in the crystal.
  Because the CO$_2$ molecule is only fitted to the cavities in the S-I form,
  the solubility of CO$_2$ becomes high.
  In addition, it was found that
  tri-atomic molecules can more effectively diffuse in the S-I form 
  compared with di-atomic molecules.
  This peculiar behavior, larger molecule can diffuse more quickly, 
  was proved by using a ``long-gas'' model, 
  in which long-axis lengths of N$_2$ and O$_2$ were extended to that of CO$_2$.
  A novel design criterion for the CO$_2$ separation membranes is proposed
  based on the aspect of momentum transfer from polymer matrix to the penetrant,
  whose re-orientational motion is coupled to jump trials between cavities.
\end{abstract}


\end{frontmatter}


\section{Introduction}

Membrane separation is expected to be widely used
because it is an energy saving and maintenance free process.
To install the membrane separation system in large-scale plants, 
such as carbon capture in power plants or natural gas fields, 
both selectivity and permeability should be raised.
However, as argued by Robeson \cite{Robeson2008} 
and reviewed by Park et. al \cite{Park2017}, 
there is a trade-off between permeability and selectivity.
In the plot of selectivity vs. permeability (so-called Robeson plot), 
an upper bound is drawn as a linear declined line.
Many efforts have been made to overcome the upper bound \cite{Park2017}.

To raise both of the competing properties, selectivity and permeability, 
we attempt to apply a ``molecular cavity'' 
to porous polymer crystal \cite{Tamai2003b}.
The molecular cavity is a small vacant space in crystals, 
in which a guest molecule can be just fitted.
The cavity is characterized by 
regular size, regular shape, and ordered connectivity.
The permeability of gases in the molecular cavity may be controlled 
by external stimulus.
For example, by applying stress to the porous crystal, 
the connectivity of molecular cavities may be changed 
and the size and connectivity of permeation channels may be controlled; 
penetrants are precisely separated by size, 
and only particular species can be effectively permeated.

The use of ordered structures to the separation processes is familiar 
in the field of inorganic membrane, 
such as zeolites \cite{Liu2021,Hasanzadeh2021}, 
graphene \cite{Yuan2017,Yuan2019}, and
various nano-sheets \cite{Bayat2020,Azamat2018}.
One of the advantages of using porous polymer crystals is that 
the main chains are rather flexible.
Local segmental motion of the chains may help the transport of gas molecules.

Some crystalline forms of syndiotactic polystyrene (s-PS) are 
expected to be candidates to realize the concept of the molecular cavity,
because of their porous nature \cite{Milano2009,Guerra2012}.
As reviewed by some authors \cite{Gowd2009,Milano2009,Guerra2012}, 
s-PS is known for complex polymorphic behavior.
There are five main polymorphs: 
$\alpha$ \cite{Greis1989,DeRosa1991,DeRosa1996a,Cartier1998}, 
$\beta$  \cite{DeRosa1992,Chatani1993a}, 
$\gamma$ \cite{Wang1992,Rizzo2002,Tamai2002b,Tamai2018}, 
$\delta$ \cite{Chatani1993,Chatani1993b,DeRosa1999,DeRosa1997}, and 
$\varepsilon$ \cite{Rizzo2007,Petraccone2008} form.
The $\delta$ and $\varepsilon$ forms have porous nature; 
various guest molecules are clathrated in these forms 
\cite{Tarallo2010a,Tarallo2011a,Acocella2015}.
The re-orientational dynamics of guests are significantly affected 
by host--guest interaction, depending on the guest structure and cavity shape
\cite{Tamai2005,Kobayashi2016,Kobayashi2018}.
The diffusion of small molecules in these crystals is anisotropic,
reflecting crystalline orientations \cite{Milano2002,Tamai2003b,Venditto2006}.
Narrow channels are also exist in the $\alpha$ form, 
in which gas molecules are permeable \cite{Hodge2001,Prodpran2002,Tamai2003a},
while more dense $\beta$ form is impermeable 
to small gas molecules \cite{Prodpran2002}.
The diffusion coefficients of gases in the porous $\delta$ form are
about two order of magnitude lower than that in the amorphous phase 
\cite{Larobina2004}, because each cavity is separated in this form.

These crystalline forms are interchangeable with each other
by thermal or solvent treatments \cite{Gowd2009}.
External stresses affect the transition \cite{DeCandia1991,Ouchi2011,Endo2018}.
The polymorphic behavior of s-PS were investigated also 
by atomistic molecular dynamics (MD) simulations
\cite{Tamai2002b,Tamai2017,Tamai2018,Liu2018,Liu2020}
and coarse-grained simulations \cite{Liu2018,Liu2020}.

Recently, the author found remarkable structural transition behavior
of a subnano-porous crystal of s-PS 
by using a molecular dynamics (MD) simulation \cite{Tamai2013}.
A new transition path, 
$\varepsilon$ $\rightarrow$ S-I $\rightarrow$ $\gamma$ form, 
by a uniaxial stress treatment was found, 
where the S-I form (1st form under stress) was only stable under the stress.
The initial and end points, $\varepsilon$ and $\gamma$ forms, respectively, 
are experimentally known forms, while the S-I is a newly found form.
The rearrangement of nano-porous cavity structures 
associated with stress-induced phase transitions 
was also reported \cite{Tamai2013}.
It was demonstrated that the nano-scale channel in the $\varepsilon$ form 
was transformed into a zigzag channel suitable for gas separations. 
From preliminary simulations \cite{Tamai2013Pa}, 
the permeability of gases in the S-I form exhibited the possibility of 
effective CO$_2$/N$_2$ and CO$_2$/CH$_4$ separation.

The reproducibility of the transition and stability of the S-I form 
were thoroughly confirmed in a previous study \cite{Tamai2017}.
Various thermal and stress induced transitions of s-PS were examined 
in the wide range of temperature--stress plane and 
a kind of phase diagram was derived.
It was found that the S-I form was 
more stable than the $\delta_{\mathrm{e}}$ and $\varepsilon$ forms and
most stable under the stress of $\sigma_{yy}$ = 0.27 GPa at 300 K.

In the present paper, 
the S-I crystal structure is fully characterized for the first time, 
using the atom coordinates obtained in the previous study \cite{Tamai2017};
the fractional coordinates of atoms in an asymmetric unit were determined.
The main objective of the present paper is investigating
permeation behaviors of gas molecules 
in the newly determined S-I form crystal by MD simulations.
The solubilities and diffusion constants of gases,
H$_2$, O$_2$, N$_2$, CH$_4$, and CO$_2$, in the S-I form are examined 
by long-time MD simulations, compared with those in the $\varepsilon$ form.
It is shown that the S-I form exhibits 
extremely high CO$_2$/N$_2$ and CO$_2$/CH$_4$ selectivity 
while preserving its high CO$_2$ permeability.
A noteworthy finding is that 
tri-atomic molecules can more effectively diffuse in the S-I form
compared with di-atomic molecules.
In other words, a larger molecule can diffuse more quickly.
In order to clarify the mechanism of this peculiar behavior, 
further MD simulations are performed using a ``long-gas'' model,
in which long-axis length of N$_2$ and O$_2$ is extended to that of CO$_2$.
The results are interpreted in terms of effective momentum transfer 
from the polymer matrix to the penetrants.

\section{Simulation details}
\subsection{Force field}

The force field for s-PS, AMBER \cite{Cornell1995}, 
was
the same as those used previously \cite{Tamai2013,Tamai2017,Tamai2018}.
The applicability of the force field
was confirmed in the previous papers.
The partial charges were assigned only for aromatic hydrogens ($+0.085$e) and
aromatic carbons bonded to the hydrogen ($-0.085$e).
Using the model, experimentally determined crystal structures of 
s-PS polymorphs were satisfactorily reproduced.
Various transition paths among these polymorphs 
by temperature or stress treatments 
were also reproduced by the model \cite{Tamai2017,Tamai2018}.

\begin{table}[ht]
  \caption{
    Force-field parameters for gases 
    \cite{Potoff2001,Hansen2007,Yang2006,Kaminski1994}.
  }
  \label{tbl:FF-parm}

  \begin{center}
  \begin{tabular}{ccllll}
    \hline
    & 
    & \multicolumn{1}{c}{$\sigma_{\mathrm{LJ}}$}
    & \multicolumn{1}{c}{$\varepsilon_{\mathrm{LJ}}$}
    & \multicolumn{1}{c}{$q$}
    & \multicolumn{1}{c}{$l_{\textrm{b}}$\textsuperscript{\emph{b}}}
    \\
      \raisebox{1ex}[0pt]{gas} 
    & \raisebox{1ex}[0pt]{atom\textsuperscript{\emph{a}}}
    & \multicolumn{1}{c}{({\AA})}
    & \multicolumn{1}{c}{(kJ/mol)}
    & \multicolumn{1}{c}{(e)}
    & \multicolumn{1}{c}{({\AA})}
    \\
    \hline
    CO$_2$ & O & 3.05  & 0.657 & $-$0.35  & 1.16  \\
           & C & 2.80  & 0.224 & $+$0.70  & \\
    N$_2$  & N & 3.31  & 0.299 & $-$0.482 & 0.55  \\
           & M & 0     & 0     & $+$0.964 & \\
    O$_2$  & O & 3.013 & 0.408 & $-$0.123 & 0.605 \\
           & M & 0     & 0     & $+$0.246 & \\
    H$_2$  & H & 0     & 0     & $+$0.468 & 0.37  \\
           & M & 2.958 & 0.305 & $-$0.936 & \\
    CH$_4$ & H & 2.50  & 0.126 & $+$0.060 & 1.09  \\
           & C & 3.50  & 0.276 & $-$0.240 & \\
    \hline
  \end{tabular}
  \end{center}

  \raggedright
  \textsuperscript{\emph{a}}M is massless point to reproduce quadrupole.\\
  \textsuperscript{\emph{b}}Bond length to the center of mass of molecule.
\end{table}

\begin{figure}[ht]
  \centering
  \includegraphics[width=82mm]{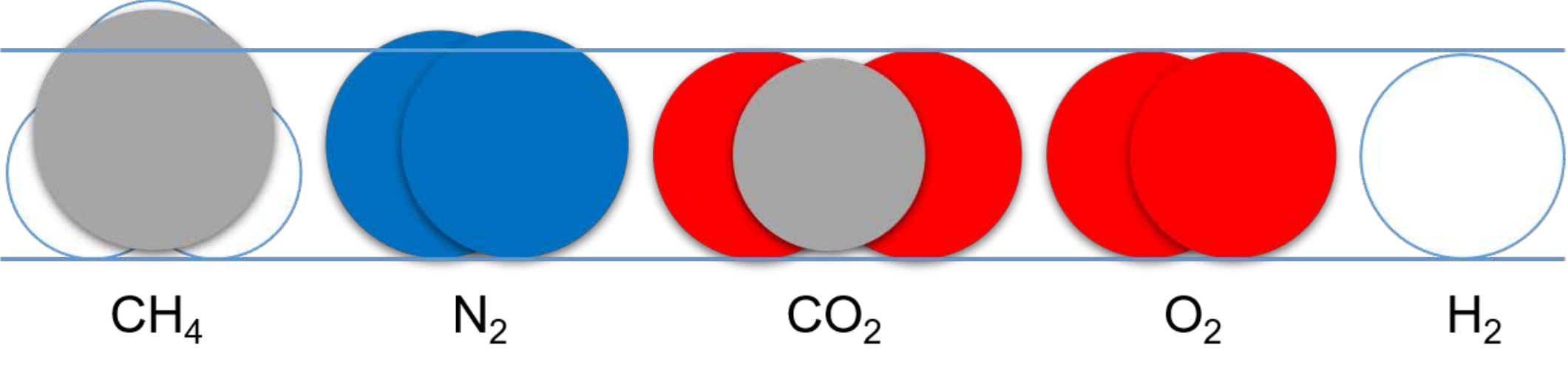}

  \caption{Schematic representation of gas models, 
    based on $l_{\textrm{b}}$ and $\sigma_{\mathrm{LJ}}$.}
  \label{fig:gas-model}
\end{figure}

The force-field parameters for gas molecules are listed 
in Table \ref{tbl:FF-parm},
where $\sigma_{\mathrm{LJ}}$ and $\varepsilon_{\mathrm{LJ}}$ are the size and 
energy parameters, respectively, of 12--6 Lennard-Jones (LJ) potential 
and $q$ is the partial charge.
The CO$_2$, N$_2$, O$_2$, and H$_2$ molecules were modeled as
rigid body with quadrupole.
The quadrupole moments were modeled by three partial charges; 
an additional interaction site, M, was introduced 
at the center of each diatomic molecule.
The transferable potential for phase equilibria (TraPPE) force field
developed by Potoff and Siepmann \cite{Potoff2001}
was used for CO$_2$ and N$_2$.
The model can quantitatively reproduce the vapor-liquid equilibria 
of neat N$_2$ and CO$_2$ and also their mixtures with alkanes \cite{Potoff2001}.
The TraPPE force field with quadrupole was also used 
for O$_2$ \cite{Hansen2007}.
The model by Yang et al.\cite{Yang2006} was used for H$_2$, 
in which a single LJ site was located between two H atoms 
and three partial charges were settled on the molecule.
OPLS-AA \cite{Kaminski1994} was used for CH$_4$.
The schematic representation of gas models are 
shown in Fig. \ref{fig:gas-model}, 
in which sphere size reflects the LJ size parameter $\sigma_{\mathrm{LJ}}$.

These gas models have been used extensively 
in combination with other force fields such as AMBER, OPLS-AA, etc.
to investigate gas permeation behavior through membranes.
For example, the TraPPE was used to simulate permeation behavior of 
CO$_2$ and N$_2$ through graphene membranes \cite{Yuan2017,Yuan2019}, 
in combination with AMBER and OPLS-AA force fields.
The model of Yang et al. was used to simulate adsorption of CO$_2$/CH$_4$/H$_2$
mixture in metal--organic frameworks (MOFs) \cite{Yang2006},
associated with TraPPE and OPLS-AA.

In our previous study \cite{Tamai2003a,Tamai2003b,Asahi2015}, 
sorption and diffusion of CO$_2$, O$_2$, and N$_2$ 
in s-PS $\alpha$ and $\delta_{\mathrm{e}}$ forms
have been investigated by other classical gas models: 
MSM for CO$_2$ \cite{Murthy1981} and two center LJ model 
(without quadrupole) for O$_2$ and N$_2$ \cite{Fischer1983,Cheung1975}.
The s-PS chains were modeled by AMBER, the same model as in the present study.
The same set of force fields was also applied for CO$_2$ sorption 
in the $\delta_{\mathrm{e}}$ form by other authors \cite{Sanguigno2011}; 
the sorption isotherms of CO$_2$ were successfully reproduced.

To confirm the sensitivity of the force fields on gas permeation behavior,
I have also performed 
gas-permeation simulation in the s-PS S-I and $\varepsilon$ forms
using the latter set of the force fields
in addition to the parameter set tabulated in Table \ref{tbl:FF-parm}.
The results are summarized in Tables S2--S5 of Supporting Information.
It was found that conclusions of the present paper 
were not affected by the force fields used.
From these reasons, 
the use of the present force field (Table \ref{tbl:FF-parm}) is justified.

\subsection{Molecular dynamics simulation}
The method of MD simulation was the same as
those used previously \cite{Tamai2013,Tamai2017,Tamai2018}.
Three-dimensional periodic boundary condition (PBC) was applied, i.e., 
infinitely spread single crystals were simulated.
Each polymer chain was connected infinitely by the PBC.
The LJ interaction was smoothly cut off at 12 {\AA} 
with a long range correction.
The long-range Coulomb interactions were handled by the Ewald sum method.
The Nos\'e \cite{Nose1984} method was used to control temperature (300 K) and
the Parrinello--Rahman \cite{Parrinello1981} method was used 
to control the pressure tensor of the system.
All the bond lengths of polymer and CH$_4$ were constrained 
by the SHAKE method\cite{Ryckaert1977}.
The other gas molecules were treated as rigid body 
by the SHAKE vectorial constraint method \cite{Ryckaert1977,Ciccotti1982}.
The equations of motion were solved by a variant of the Verlet 
algorithm \cite{Verlet1967,Ferrario1985}, with a time step of 1 fs.
Although no constraint forces to keep the center of mass (COM) of whole crystal
were applied, drifts of the COM were negligible.

\subsection{Crystal-structure determination}

The s-PS S-I form was obtained in the previous study, in which
the initial $\varepsilon$ form crystal was transformed into the S-I form
by stress treatment \cite{Tamai2017}.
Specifically,
a uniaxial stress along the $b$-axis was gradually applied to the crystal 
by changing the $yy$ component of the symmetric tensor $\mathbf{\Sigma}$
stepwise at a rate of 1 kJ/mol{\AA} per 130 ps.
The symmetric tensor $\mathbf{\Sigma}$ of 
the Parrinello--Rahman method \cite{Parrinello1981}
is related to the external stress $\mathbf{\sigma}$ as: 
\begin{equation}
  \mathbf{\Sigma} 
  = {\mathbf{h}_0}^{-1} ( \mathbf{\sigma} - p ) ({\mathbf{h}_0}^{-1})^{{t}} V_0, 
  \label{eqn:Stress}
\end{equation}
where $\mathbf{h}$ is the cell matrix, $p$ is hydrostatic pressure, 
$V$ is the volume of the MD unit cell, and
the subscript 0 denotes the reference state.
Starting from the $\varepsilon$ form with 3$\times$2$\times$6 crystal units
the structural transition to the S-I form was initiated 
at $\Sigma_{yy}$ = 16 kJ/mol{\AA} (at 300 K), 
which corresponded to a stress of $\sigma_{yy}$ = 0.43 GPa.
The stress was released to 0.27 GPa through the structural transition.
After further compression to $\Sigma_{yy}$ = 25 kJ/mol{\AA}
($\sigma_{yy}$ = 0.41 GPa), 
the stress was gradually reduced to $\Sigma_{yy}$ = 15 kJ/mol{\AA}
($\sigma_{yy}$ = 0.27 GPa).
In the previous study \cite{Tamai2017}, 
the thermal stability of the S-I form was examined 
under various $\Sigma_{yy}$ values.
Because the S-I form was found to be the most stable 
under a stress of $\sigma_{yy}$ = 0.27 GPa, 
that structure was examined in this study.

The structure factor $F_{hkl}$ and X-ray intensity $I_{hkl}$ 
for the Miller index $(hkl)$ was calculated by a procedure shown in
\ref{sec:intensity}.
The intensity $I_{2\theta}$ for diffraction angle $2\theta$ 
was calculated by summing $I_{hkl}$ values for $(hkl)$s 
that are related to the same $2\theta$ within a small tolerance. 
The $F_{hkl}$, $I_{hkl}$, and $I_{2\theta}$ values were 
averaged over 200 configurations during 100 ps.
The crystal space group was determined by examining 
the deduction of $F_{hkl}$ for particular sets of $hkl$.

The fractional coordinates of atoms were calculated for each coordinate 
every 500 steps (0.5 ps), and averaged over 100 ps trajectories.
The averages were also taken over all 72 crystal units in the S-I MD unit cell.
The coordinates of an asymmetric unit were selected 
from the averaged coordinates of eight monomer units.

\subsection{Simulation of gas permeation}

The S-I crystal model consisted of 4$\times$3$\times$6 crystal units 
was generated using the determined space group information 
and coordinates of the asymmetric unit.
In this MD unit cell setting, $\Sigma_{yy}$ is equal to 7 kJ/mol{\AA} 
(instead of $\Sigma_{yy}$ = 15 kJ/mol{\AA}) 
to conform $\sigma_{yy}$ = 0.27 GPa.
The condition is different from that used in the preliminary 
calculation \cite{Tamai2013Pa}, where $\sigma_{yy}$ = 0.18 GPa 
using the MD unit cell with 6$\times$2$\times$6 crystal units.
The structure of the $\varepsilon$ form 
consisted of 3$\times$2$\times$6 crystal units was generated 
based on the experimentally determined structure \cite{Petraccone2008}.
The crystals were equilibrated for 200 ps at 300 K.

The excess chemical potential $\mu_{\mathrm{r}}$ of gas molecule 
in the crystal was calculated 
by the Widom particle insertion method \cite{Widom1963} as
\begin{equation}
  \mu_{\mathrm{r}} = - k_{\mathrm{B}}T \ln 
  \langle \exp \left( - u / k_{\mathrm{B}}T \right) \rangle, 
  \label{eqn:mur}
\end{equation}
where $u$ is energy difference before and after insertion of a test molecule, 
$k_{\mathrm{B}}$ is Boltzmann constant, $T$ is temperature, and 
$\langle \cdots \rangle$ means ensemble average.
100,000 insertion trials of the test molecule were attempted to each of
200 configurations during 100 ps.
The excluded volume map sampling (EVMS)
was used to raise the efficiency of sampling \cite{Tamai1995a}.
To estimate the standard error, sampling points were divided into 
40 subsets, each of which consists of 5 configurations.
The solubility $S$ was calculated by
\begin{equation}
  S = \exp \left( - \mu_{\mathrm{r}} / k_{\mathrm{B}}T \right),
  \label{eqn:S}
\end{equation}
whose unit was translated into cm$^3$(STP)/cm$^3$atm
by multiplying $273.15/T$, where $T$ is temperature in K (300 K).

In order to simulate the diffusion behavior, 
gas molecules were inserted into the crystal,
such that only one molecule diffuse in each separated channel.
A total of 24 and 12 molecules were inserted in the S-I and $\varepsilon$ form,
respectively.
The simulation runs of 10 ns were performed separately for 
five gas species, H$_2$, O$_2$, N$_2$, CH$_4$, and CO$_2$, at 300 K.
The simulations
of gas diffusion
were repeated for three times for each system 
and obtained physical quantities were averaged over three runs.

All simulation runs and numerical analyses were performed with 
the PAMPS \cite{Tamai1994} molecular simulation program coded by the author,
using the super computer at ACCMS, Kyoto University.

\section{Results and discussion}
\subsection{Crystal structure of S-I form}

\begin{figure*}[ht]
  \includegraphics{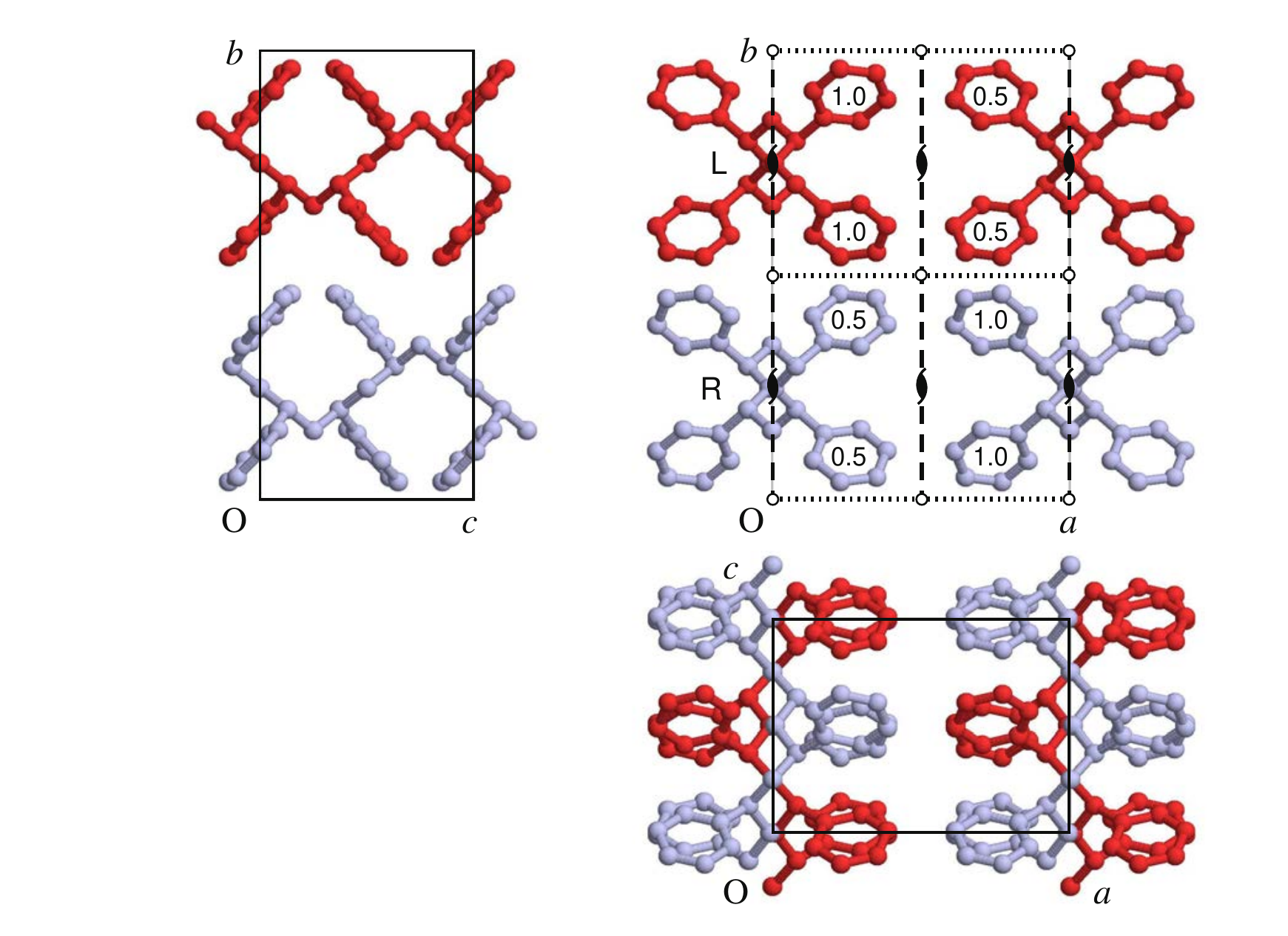}

  \caption{Snapshots of a s-PS S-I form crystal 
    under the stress of $\sigma_{yy}$ = 0.27 GPa.
    The right-handed (R) and left-handed (L) helices are 
    light blue and red, respectively.
    The approximate fractional coordinates $z/c$ of the barycenters 
    of the phenyl rings are shown.
  }
  \label{fig:snap-S-I}
\end{figure*}

A snapshot of the S-I and $\varepsilon$ forms are shown 
in the previous paper \cite{Tamai2017}.
The crystal system of these forms is orthorhombic.
The symmetry of the crystal was changed by the structural transition, and
one crystal unit of the $\varepsilon$ form was separated into 
two units in the S-I form.
In the crystal unit, 
there are eight monomer units (two helices) in the S-I form, 
whereas there are 
sixteen monomer units (four helices) in the $\varepsilon$ form.
A crystal unit of the S-I form under $\sigma_{yy}$ = 0.27 GPa stress 
is depicted in Fig. \ref{fig:snap-S-I}.
One right-handed (R, blue) and one left-handed (L, red) helix
are contained in a unit cell.
Some monomer units are translationally duplicated for better visualization.
The crystal lattice parameters of the S-I form are listed 
in Table \ref{tbl:crystal}, compared to the other forms.
Note that the lattice parameters $a$ and $b$ vary with 
external stress $\sigma_{yy}$;
the values of $a$ and $b$ are reduced or enlarged, respectively, 
with decreasing $\sigma_{yy}$.
The S-I form can exist above $\sigma_{yy}$ = 0.04 GPa in the condition of 
the previous study \cite{Tamai2017}, including a metastable state.
The density of the S-I form, 0.969 g/cm$^3$, is approximately the same as 
that of the empty $\delta_{\mathrm{e}}$ form.

\begin{table*}[ht]
  \caption{Lattice constants of s-PS crystalline polymorphs.}
  \label{tbl:crystal}

  \begin{center}
  \begin{tabular}{llccccc}
    \hline
    form & space group 
    & $a$ (\AA) & $b$ (\AA) & $c$ (\AA) & $\gamma$ (deg.) & $\rho$ (g/cm$^3$)\\
    \hline
    S-I\textsuperscript{\emph{a}}
    & $Pbcb$   &  10.96 &  16.55 &  7.87 &   90~  & 0.969 \\
    $\varepsilon$\textsuperscript{\emph{b}}
    & $Pbcn$   &  16.1~ &  21.8~ &  7.9~ &   90~  & 0.98~ \\
    $\gamma$-I\textsuperscript{\emph{c}}
    & $P2_1/a$ &  19.42 &  ~8.52 &  7.93 &  ~83.4 & 1.061 \\
    $\gamma$-II\textsuperscript{\emph{c}}
    & $I2$     &  19.37 &  17.26 &  7.76 &  ~96.4 & 1.073 \\
    $\delta_{\mathrm{e}}$\textsuperscript{\emph{d}}
    & $P2_1/a$ &  17.4~ &  11.85 &  7.70 &  117~~ & 0.977 \\
    \hline
  \end{tabular}
  \end{center}

  \raggedright
  \textsuperscript{\emph{a}}This work, under stress of $\sigma_{yy}$ = 0.27 GPa.\\
  \textsuperscript{\emph{b}}Petraccone et al. \cite{Petraccone2008}\\
  \textsuperscript{\emph{c}}Simulation by Tamai \cite{Tamai2018}.\\
  \textsuperscript{\emph{d}}De Rosa et al. \cite{DeRosa1997}, empty form.
\end{table*}

\begin{table}[ht]
  \caption{Fractional atomic coordinates for an asymmetric unit 
    of the s-PS S-I form for the space group $Pbcb$.}
  \label{tbl:ASU}

  \begin{center}
  \begin{tabular}{lcccc}
    \hline
    atom\textsuperscript{\emph{a}} &  $x/a$ &  $y/b$ &  $z/c$ & occupancy \\
    \hline
    C1             &  0.008 &  0.250 &  0.500 & ~~0.5 \\
    C2             &  0.000 &  0.345 &  0.750 & ~~0.5 \\
    C$_{\alpha}$     &  0.081 &  0.298 &  0.629 & 1 \\
    C$_{\mathrm{A}}$1 &  0.175 &  0.352 &  0.548 & 1 \\
    C$_{\mathrm{A}}$2 &  0.141 &  0.411 &  0.432 & 1 \\
    C$_{\mathrm{A}}$3 &  0.228 &  0.459 &  0.354 & 1 \\
    C$_{\mathrm{A}}$4 &  0.350 &  0.449 &  0.393 & 1 \\
    C$_{\mathrm{A}}$5 &  0.385 &  0.391 &  0.509 & 1 \\
    C$_{\mathrm{A}}$6 &  0.297 &  0.343 &  0.587 & 1 \\
    \hline
  \end{tabular}
  \end{center}

  \raggedright
  \textsuperscript{\emph{a}}C1, C2: methylene carbon, 
  C$_{\alpha}$: $\alpha$ carbon, 
  C$_{\mathrm{A}}$: aromatic carbon.
\end{table}

The calculated structure factors and X-ray diffraction intensities 
are listed in Table S1 of Supporting Information.
The intensities are also plotted \textit{vs.} $2\theta$ 
in Fig. S1 of Supporting Information.
The most significant peak for the S-I form is 
around $2\theta$ = 14.9$^{\circ}$, which is attributed to the \{111\} plane.
The peak around $2\theta$ = 8$^{\circ}$ is from the (100) plane; 
the Bragg distance $d$ = 10.96 {\AA} agrees well with the $a$-length.
Systematic absences of 
$00l$ and $h0l$ reflections with $l = 2n + 1$, and absences of
$0k0$, $hk0$, and $0kl$ reflections with $k = 2n + 1$, were found. 
Those intensities had 
less than 1 ppm of the maximum diffraction intensity.
Because these observations are characteristic 
of the space group $Pbcb$ (No. 54) \cite{ITC-Vol.A},
the space group of the S-I form was determined as $Pbcb$.
The space group can be transformed into $Pcaa$ 
by changing the cell setting to $\mathbf{abc} \rightarrow \mathbf{ba\bar{c}}$.
$Pcaa$ is the same space group as that of other helical syndiotactic polymers, 
such as the clathrate form of syndiotactic poly(\textit{m}-methylstyrene) 
(s-PMMS) containing CS$_2$ \cite{Petraccone2003}.

The fractional coordinates of atoms of an asymmetric unit of the S-I form
are listed in Table \ref{tbl:ASU}.
One monomer unit plus one main-chain carbon atom constitute the asymmetric unit.
Note that the occupancy factor is 0.5 for the two main-chain carbons C1 and C2 
(other than the $\alpha$-carbon).
Fig. \ref{fig:snap-S-I} was drawn based on these fractional coordinates 
and the transformation matrices according to the space group $Pbcb$.
Graphical symbols for the symmetry elements \cite{ITC-Vol.A} are also given.
The main-chain C1 resides approximately on the two-fold screw axis.
There is an 'axial' glide plane (indicated by dashed lines) 
along the $b$-axis, which is a combination of $\mathbf{b}/2$ translation 
and mirror operation by the plane.
The L helix (red) can be generated from the R helix (blue) 
by this symmetry operation.

\begin{figure*}[ht]
  \includegraphics{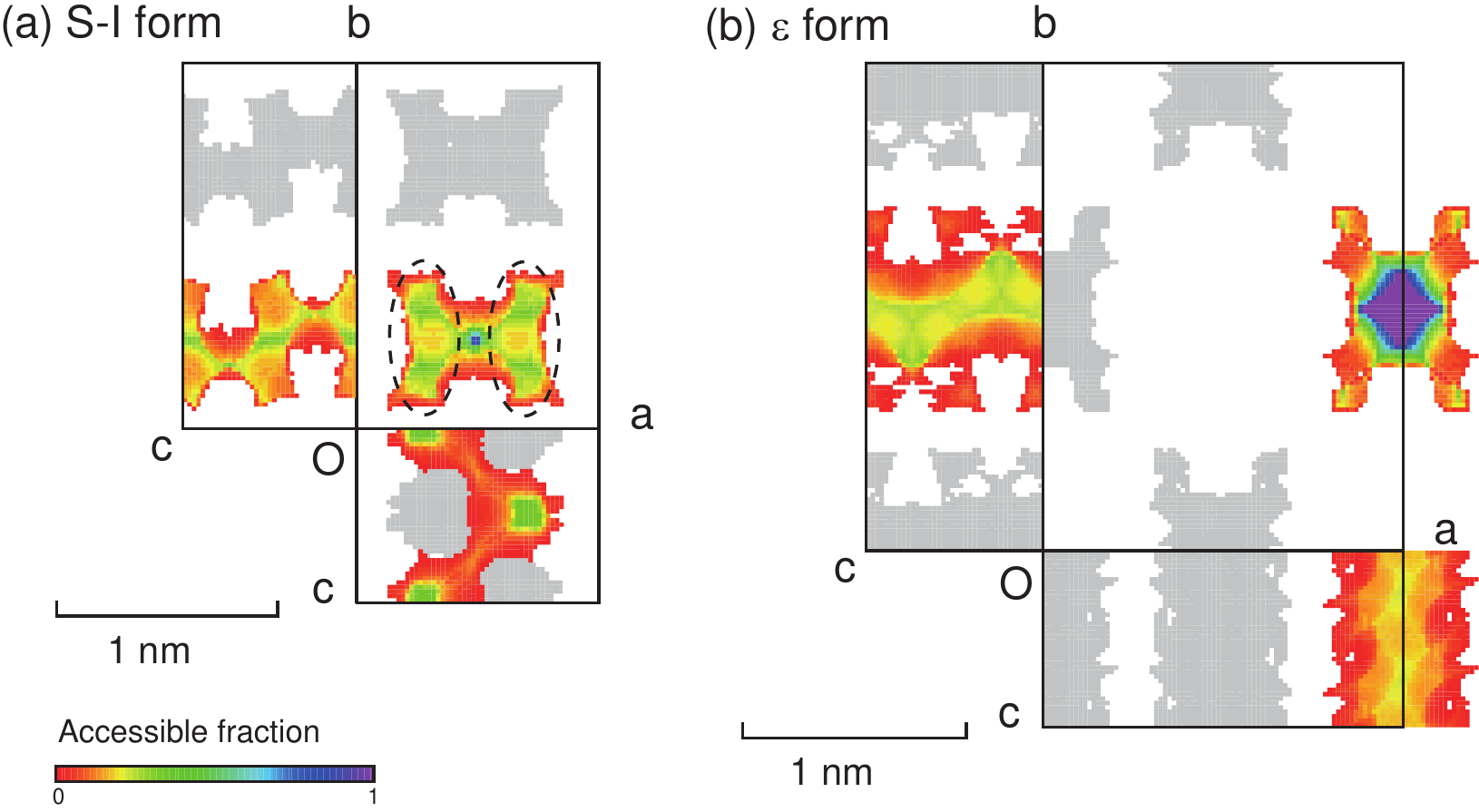}\\

  \caption{Accessible volume cluster for a probe of radius 0.7 {\AA} 
    in static structure of a crystal unit cell, 
    viewed in $a$-$b$, $b$-$c$, and $c$-$a$ planes.
    Only one continuous cluster is colored for each form for clarity.
    The color scale shows accessible fraction 
    along the normal axis to the plane;
    ``1'' means completely accessible along the normal axis,
    i.e., a straight channel exists.
    Small ellipsoidal cavities in the S-I form were indicated by dashed lines.
  }
  \label{fig:fv-cluster}
\end{figure*}

\subsection{Cavity structure}

The free volume clusters in the S-I and $\varepsilon$ forms were analyzed.
The cell was divided into a three-dimensional grid 
with an interval of 0.2 {\AA} and the grid points 
accessible for spherical probes of radius $R$ were extracted.
A cluster analysis \cite{Tamai1995a} was also performed 
for the accessible grid points.

Fig. \ref{fig:fv-cluster} shows the accessible volume cluster 
for a probe of $R$ = 0.7 {\AA}, calculated for the static 
(initial or experimental) structure of a crystal unit cell.
Only one continuous cluster is drawn for each crystal cell.
The size and position of the cavities reflect 
the chain arrangements and contacts.
In the $\varepsilon$ form, wide tube-like channels along the $c$-axis appear.
It is known that large solvent molecules, such as $n$-octane, 
can be hosted in the $\varepsilon$ form \cite{Tarallo2011a}.

In the S-I form, on the other hand, 
smaller cavities, indicated by dashed ellipse in the figure,
are alternatively connected by narrow channels; 
zigzag channels along the $c$-axis are formed.
It is expected that 
the CO$_2$ molecule is just fitted in the small cavities of the S-I form.

\begin{figure}[ht]
  \centering
  \includegraphics[width=82mm]{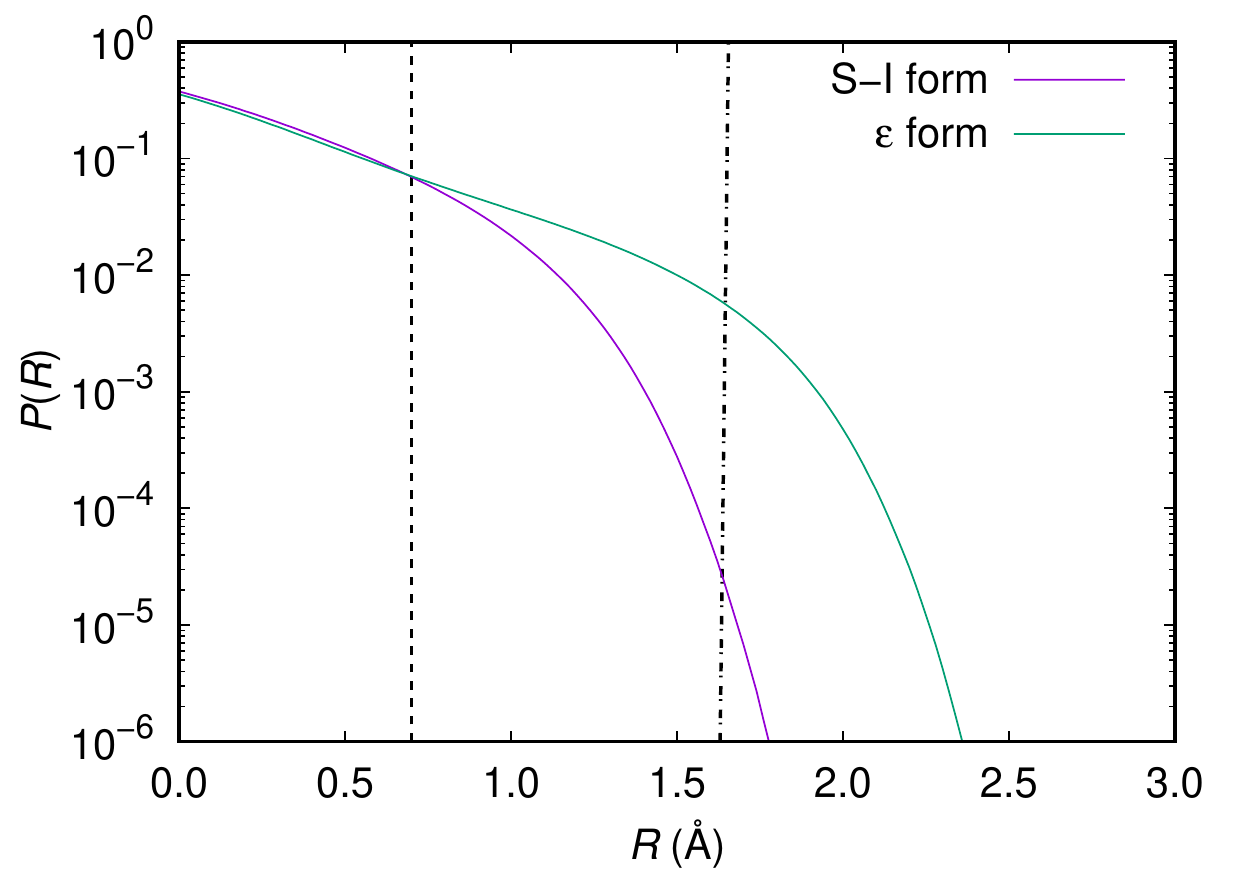}\\

  \caption{Accessible volume fraction $P(R)$ for probes of radius $R$ 
    in the S-I and $\varepsilon$ forms
    at 300 K.
    The vertical dashed line at $R$ = 0.7 {\AA} indicates the probe radius
    used for cluster analysis in Fig. \ref{fig:fv-cluster}.
    The position of $R$ = 1.655 {\AA},
    which is the half of $\sigma_{\mathrm{LJ}}$ of nitrogen atom, 
    is also shown by dash-dotted line.
  }
  \label{fig:fv-PR}
\end{figure}

Figure \ref{fig:fv-PR} shows the $R$ dependency 
of the accessible volume fraction $P(R)$ in the S-I and $\varepsilon$ forms.
The position of $R$ = 0.7 {\AA} is shown by vertical dashed line.
The effects of thermal fluctuation of the polymer chains are 
naturally incorporated because the distribution was averaged over 
200 configurations during 100 ps of MD simulation at 300 K.
The radius of nitrogen atom in N$_2$,
which is estimated as half value of $\sigma_{\mathrm{LJ}}$ for N, 
is shown by vertical dash-dotted line, 
where the accessible volume for nitrogen atom is more than two order higher
in the $\varepsilon$ form than in the S-I form.
The $P(R)$ distribution is extended to larger $R$ for the $\varepsilon$ form.
It was known that larger molecules, such as $n$-octane, can be clathrated
in the $\varepsilon$ form \cite{Tarallo2011a}.
All the gas molecules under consideration can easily access 
the $\varepsilon$ form.
On the other hand, in the S-I form, 
the slope of $P(R)$ is larger at around $R$ = 1.655 {\AA}.
Since the size of oxygen and methane is smaller and larger, respectively, 
than that of nitrogen, 
the gas molecules would be effectively filtered in the S-I form.

\subsection{Solubility of gases}

\begin{figure}[ht]
  \centering
  \includegraphics[width=82mm]{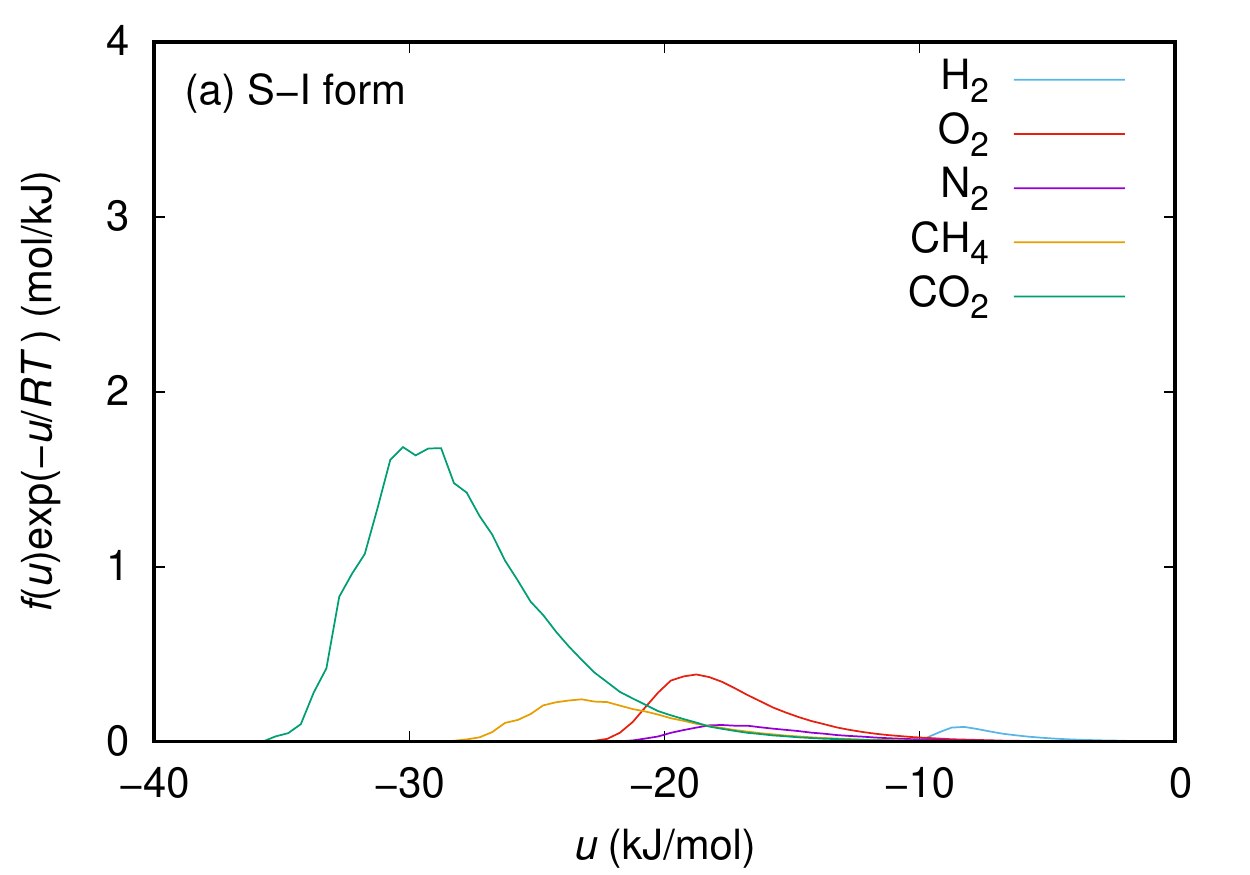}\\
  \includegraphics[width=82mm]{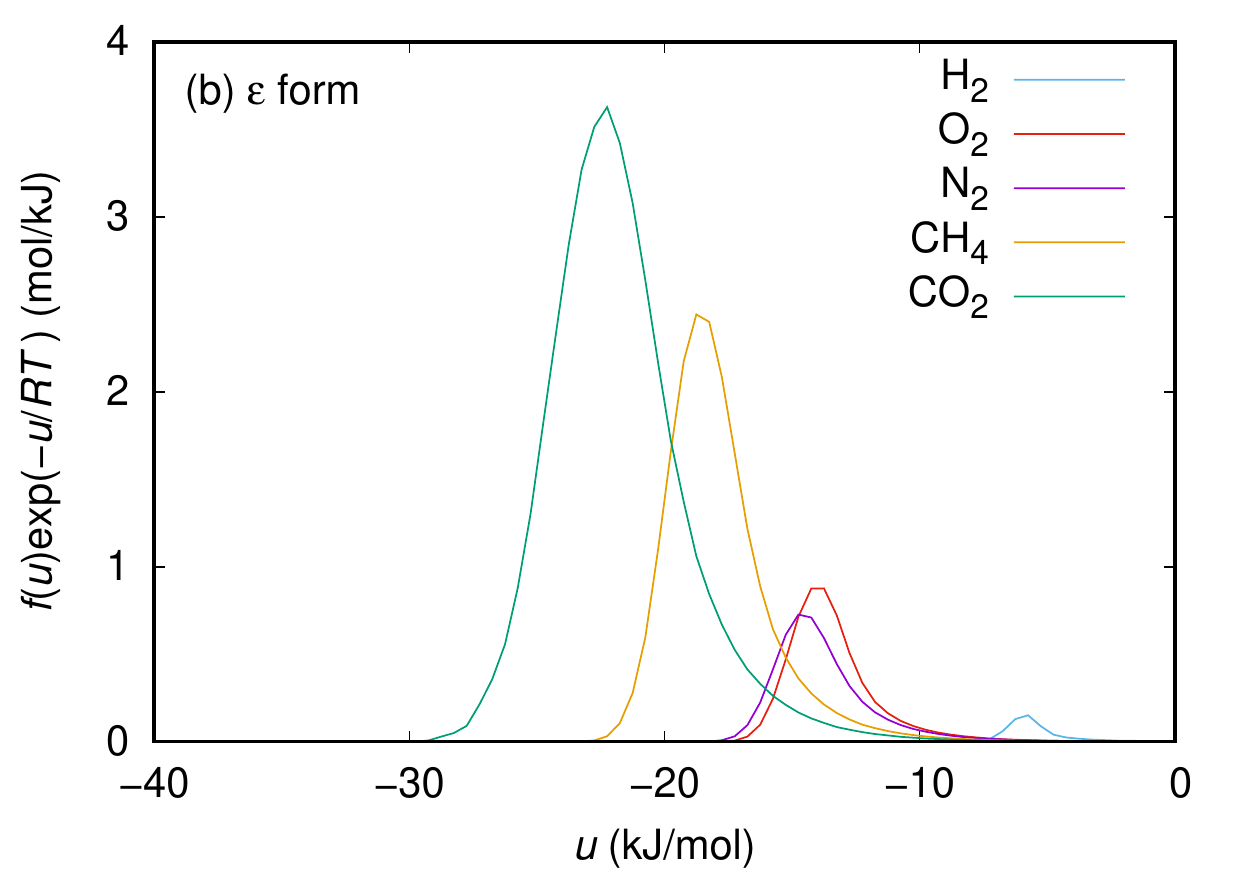}

  \caption{Plot of $f(u)\exp\left( -u / k_{\mathrm{B}}T \right)$ vs. $u$, 
    where $u$ is energy difference before and after each insertion trial.}
  \label{fig:fuexp}
\end{figure}

The solubility was calculated by the particle insertion method.
Figure \ref{fig:fuexp} shows distribution of $f(u)\exp\left( -u / RT \right)$,
where $f(u)$ is the distribution function of $u$ of insertion trials.
The value of $\mu_{\mathrm{r}}$ is related to the area of the distribution as
\begin{equation}
  \mu_{\mathrm{r}} = - k_{\mathrm{B}}T \ln
  \left[ 
    \int_{-\infty}^{\infty} f(u) \exp \left( -u / k_{\mathrm{B}}T \right) du
  \right].
  \label{eqn:mur-dist}
\end{equation}

In the $\varepsilon$ form, the distribution reflects mainly 
the attractive interaction between gas and polymer, 
rather than the repulsive interaction between atom cores, 
because the channel size is wide compared to the gas molecules,
as seen in Fig. \ref{fig:fv-cluster}.
The attractive interaction can be roughly estimated 
from the sum of LJ energy parameters, 
$\varepsilon_{\mathrm{LJ}}$ in Table \ref{tbl:FF-parm}, and in the order of 
\begin{equation}
  \mathrm{CO}_2 > \mathrm{CH}_4 > \mathrm{O}_2 > \mathrm{N}_2 > \mathrm{H}_2.
\end{equation}
Therefore, the distribution for CO$_2$ spreads over lower $u$ regions.
This leads to the slightly ragged curve for CO$_2$ in Fig. \ref{fig:fuexp}
because the curve is obtained by products of 
small $f(u)$ and large $\exp \left( -u / k_{\mathrm{B}}T \right)$
at lower energy regions.
However, the standard error of $\mu_{\mathrm{r}}$ 
is sufficiently small (Table \ref{tbl:S}).
The sampling is precise enough to evaluate $\mu_{\mathrm{r}}$ and $S$
for all the gases.

In the S-I form, the repulsive interaction, as well as the attractive one 
is important, because the cavity size is rather small.
As seen from Fig. \ref{fig:fv-cluster}, 
the CO$_2$ molecule suitably matches with the size and shape 
of the ellipsoidal cavity in the S-I form.
The three atoms of CO$_2$ reside within the attractive interaction region 
with cavity wall.
Therefore, the distribution of CO$_2$ in the S-I form 
spread over further lower $u$ regions than that in the $\varepsilon$ form.
On the other hand, 
the N$_2$ molecule is slightly fatter 
compared to the minor axis of the ellipsoidal cavity, as can be estimated 
from the LJ size parameter $\sigma_{\mathrm{LJ}}$ in Table \ref{tbl:FF-parm}.
The repulsive interaction between N$_2$ molecule and cavity wall 
suppresses the accessible volumes for N$_2$.
This leads to the lower height of the distribution for N$_2$.

\begin{table*}[ht]
  \caption{Excess chemical potential $\mu_{\mathrm{r}}$ and solubility $S$ 
    of gases in the S-I and $\varepsilon$ forms.\textsuperscript{\emph{a}}}
  \label{tbl:S}

  \begin{center}
  \begin{tabular}{cllcll}
    \hline
    & \multicolumn{2}{c}{$\mu_{\mathrm{r}}$ (kJ/mol)}
    &
    & \multicolumn{2}{c}{$S$ (cm$^3$(STP)/cm$^3$atm)}
    \\
    \cline{2-3}\cline{5-6}
    \raisebox{1ex}[0pt]{gas} 
    & \multicolumn{1}{c}{S-I form}
    & \multicolumn{1}{c}{$\varepsilon$ form}
    &
    & \multicolumn{1}{c}{S-I form}
    & \multicolumn{1}{c}{$\varepsilon$ form}
    \\
    \hline
    H$_2$  &  ~~3.0365 (0.0008) &  ~~3.1226 (0.0002) && ~0.26953   (0.00001)& ~0.26038 (0.000002)\\
    O$_2$  & $-$2.001~ (0.005)  & $-$2.641~ (0.0009) && ~2.031~~~  (0.004)  & ~2.625~~~ (0.0010) \\
    N$_2$  &  ~~1.176~ (0.012)  & $-$2.341~ (0.0012) && ~0.5682~~  (0.0007) & ~2.327~~~ (0.0011) \\
    CH$_4$ & $-$1.411~ (0.024)  & $-$5.658~ (0.0021) && ~1.603~~~  (0.011)  & ~8.80~~~~ (0.028) \\
    CO$_2$ & $-$6.437~ (0.041)  & $-$7.495~ (0.0018) && 12.0~~~~~  (1.0)    & 18.4~~~~~ (0.11)  \\
    \hline
  \end{tabular}
  \end{center}

  \raggedright
  \textsuperscript{\emph{a}}Standard errors in parenthesis.
\end{table*}

The $\mu_{\mathrm{r}}$ values, which were calculated by eq. \ref{eqn:mur},
and $S$ values are tabulated in Table \ref{tbl:S}.
In the S-I form the solubility of CO$_2$ is markedly higher than 
that of N$_2$ or CH$_4$.

\subsection{Gas diffusion in S-I form}

\begin{figure*}[ht]
  \centering
  \includegraphics[width=70mm]{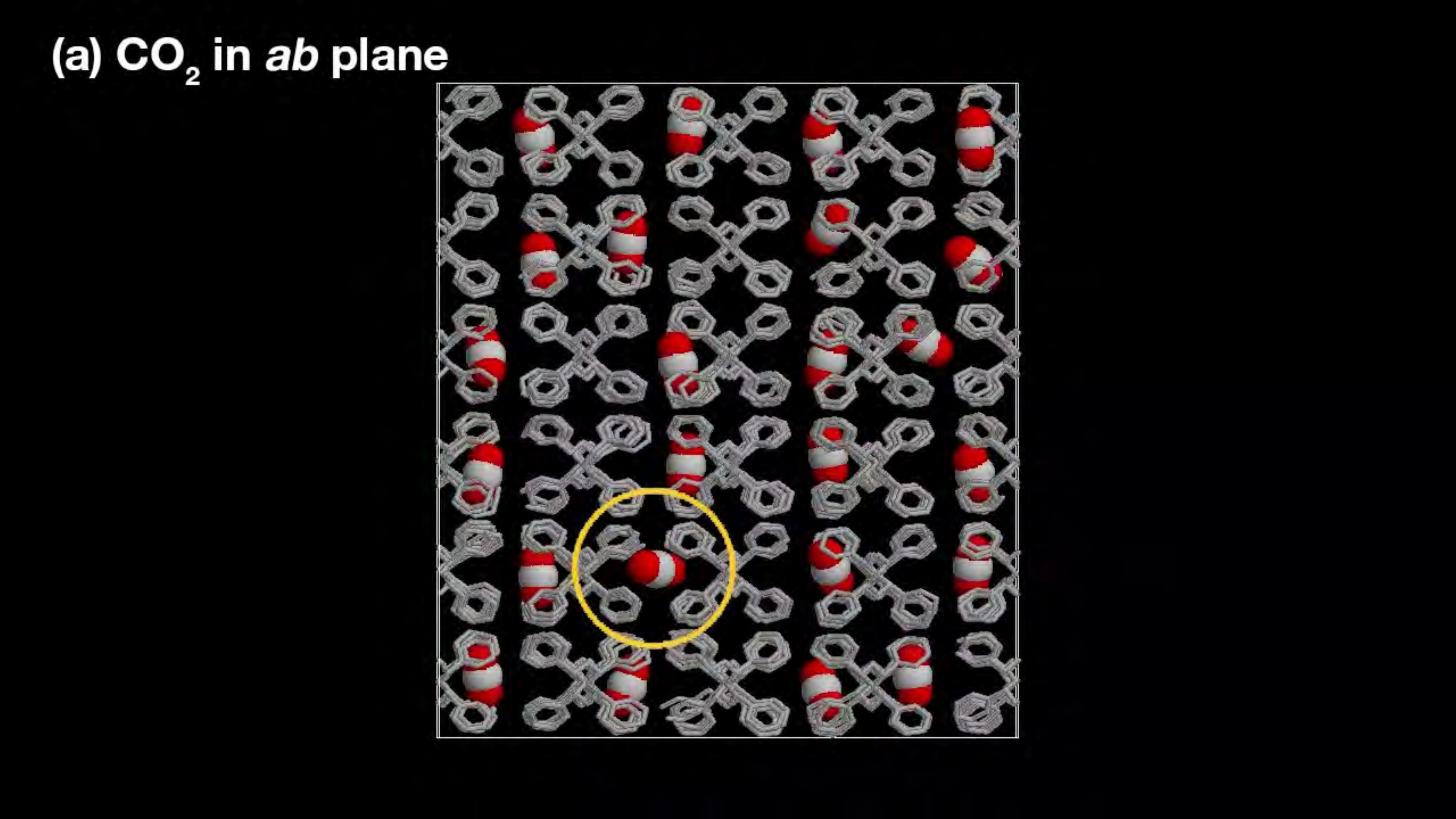}
  \includegraphics[width=70mm]{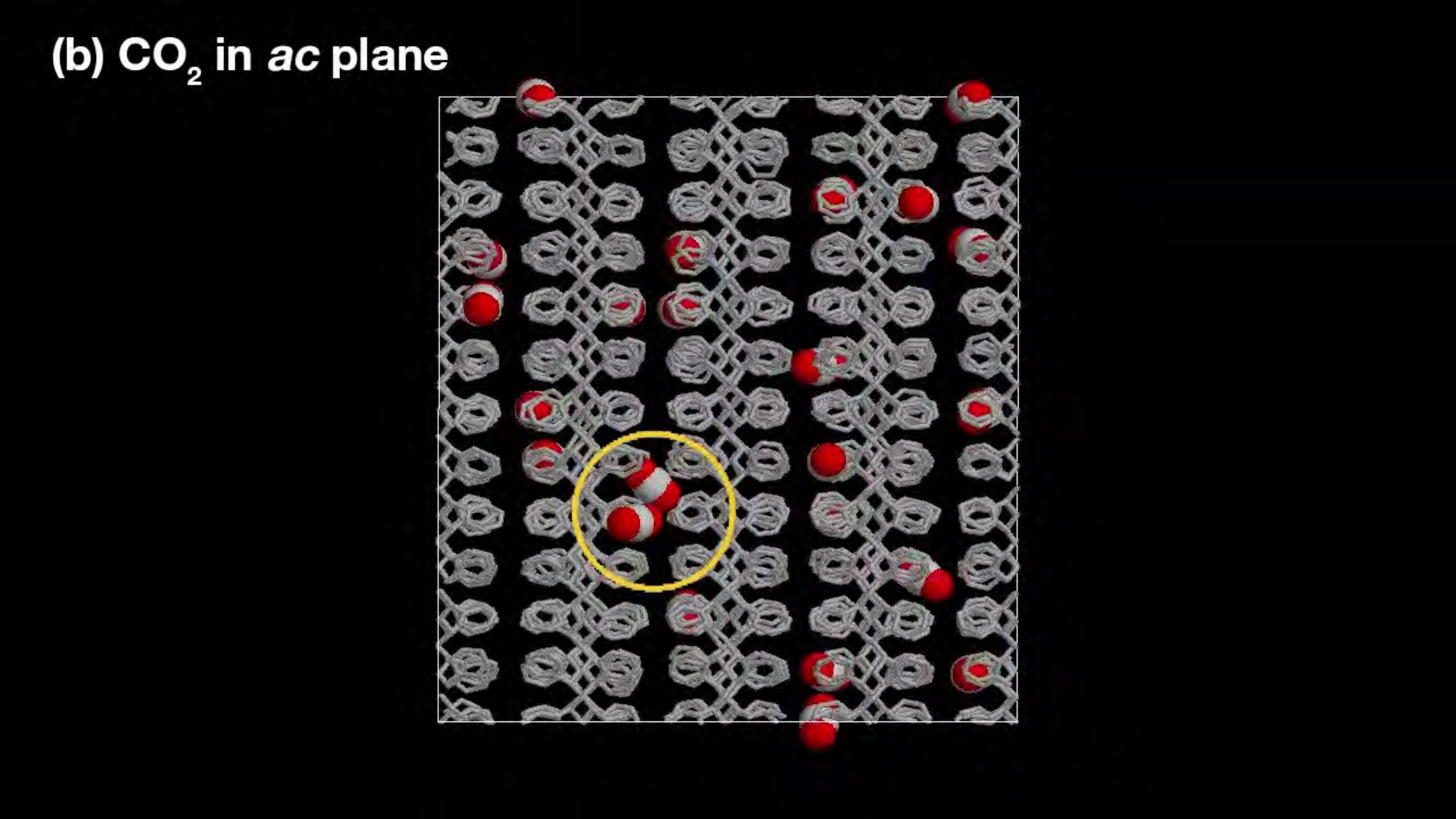}\\
  \includegraphics[width=70mm]{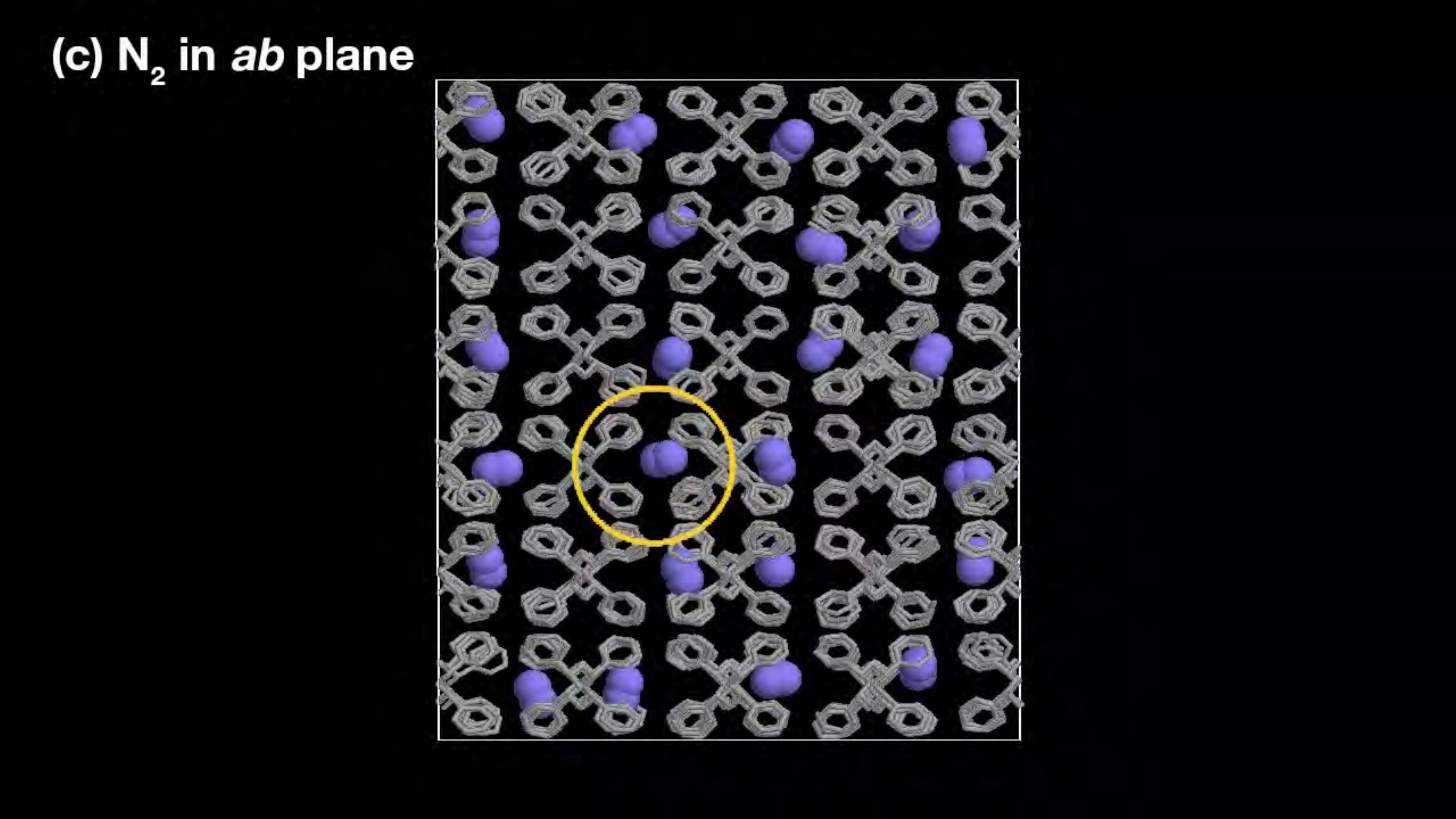}
  \includegraphics[width=70mm]{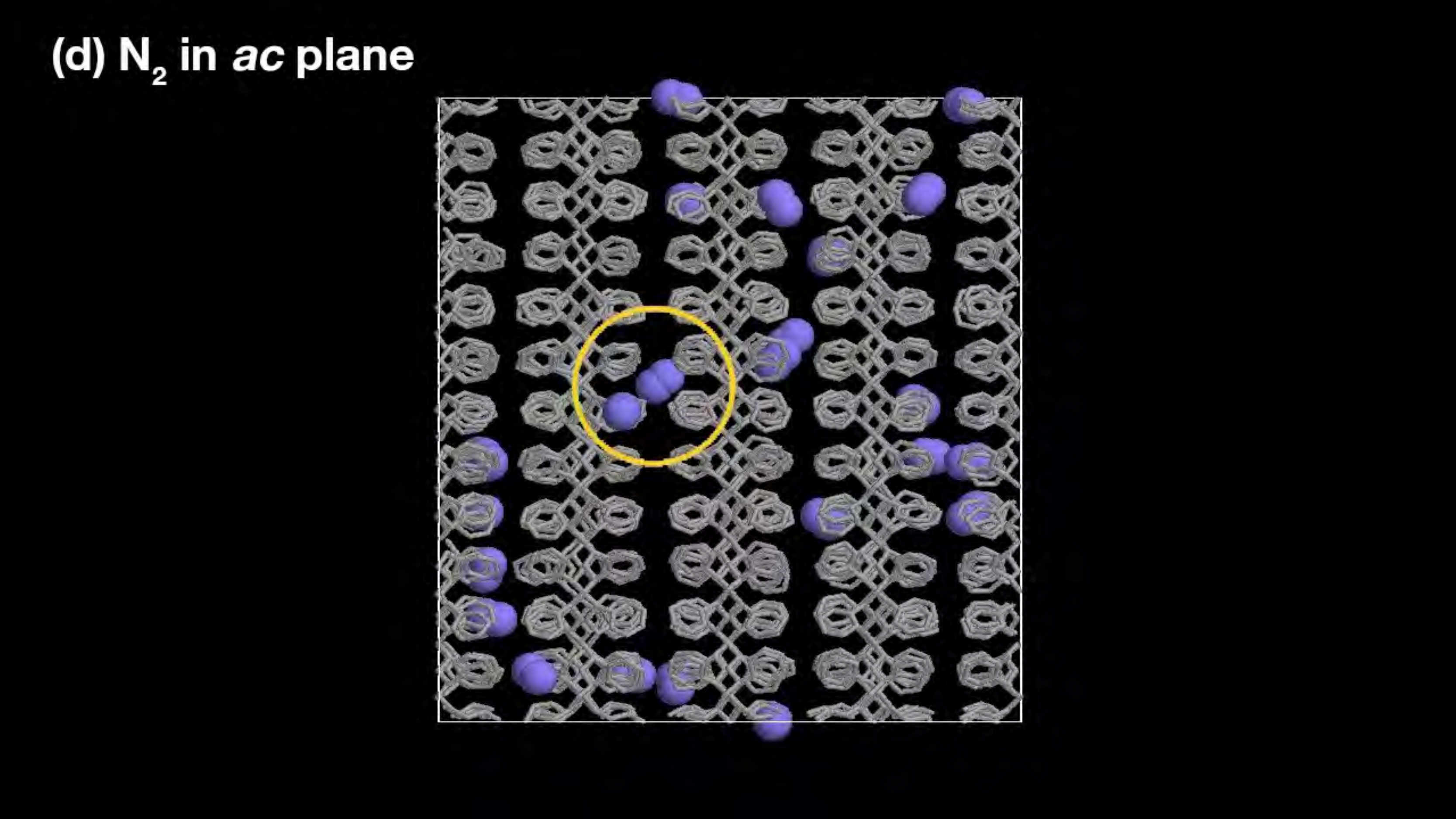}

  \caption{Snapshots at the moment of jump event of gases in s-PS S-I form.
    The pictures were extracted from movie files 
    \href{run:movie-S1-CO2-ab.mp4}{movie-S1-CO2-ab.mp4},
    \href{run:movie-S2-CO2-ac.mp4}{movie-S2-CO2-ac.mp4},
    \href{run:movie-S3-N2-ab.mp4}{movie-S3-N2-ab.mp4}, and 
    \href{run:movie-S4-N2-ac.mp4}{movie-S4-N2-ac.mp4}
    for panel (a), (b), (c), and (d), respectively.
    The jump events occur at the position indicated by yellow circles.
  }
  \label{fig:animation}
\end{figure*}

Long time simulation runs were performed 
to investigate diffusion behavior of gases in the S-I and $\varepsilon$ forms.
The animations of CO$_2$ and N$_2$ diffusion in the S-I form
viewed in the $a$-$b$ or $a$-$c$ plane are supplied as Movie S1--S4.
Snapshots at the moment of jump event of gases are also shown 
in Fig. \ref{fig:animation}, which were extracted from the movie files.

The gas molecules perform jump diffusion 
in the zigzag channels along the $c$-axis (Movie S2 and S4).
A snapshot at the moment of the jump event of CO$_2$ is also shown in Fig. S2,
where the jump path between the cavities is drawn by the arrows.
The jump probability of CO$_2$ is rather higher compared with that of N$_2$.
In the $a$-$b$ plane, 
the CO$_2$ molecules are mainly oriented along the $b$-axis and
oscillate around that orientation (Movie S1).
A perpendicular orientation is sometimes observed, where 
the CO$_2$ molecule is just jumping to the neighbor cavity
through the narrow channel (Fig. S2).
As for N$_2$, the molecules are frequently rotating in each cavity (Movie S3).
This difference in molecular motion leads to an effective transport mechanism 
of tri-atomic molecules in the S-I form, as discussed in a later subsection.

%

\begin{figure}[ht]
  \centering
  \includegraphics[width=82mm]{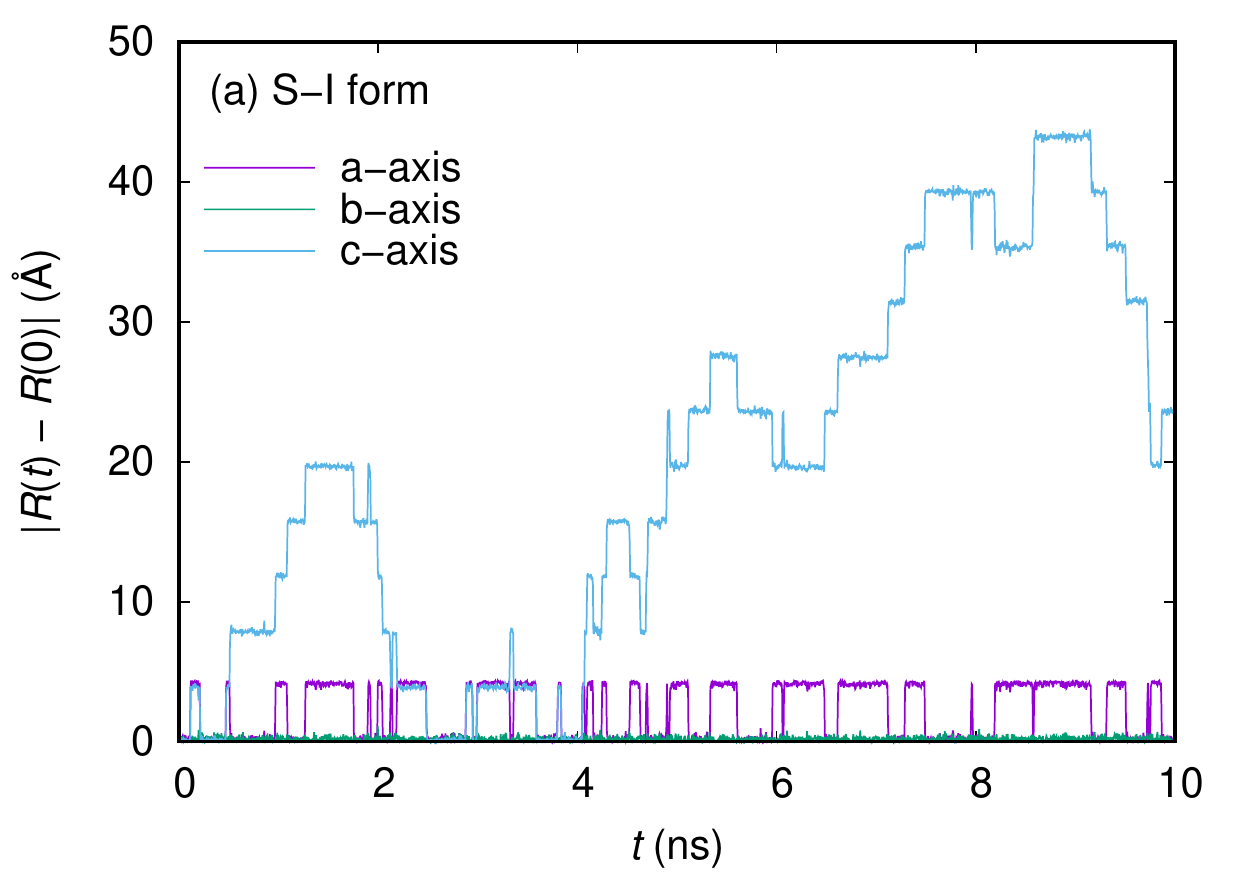}\\
  \includegraphics[width=82mm]{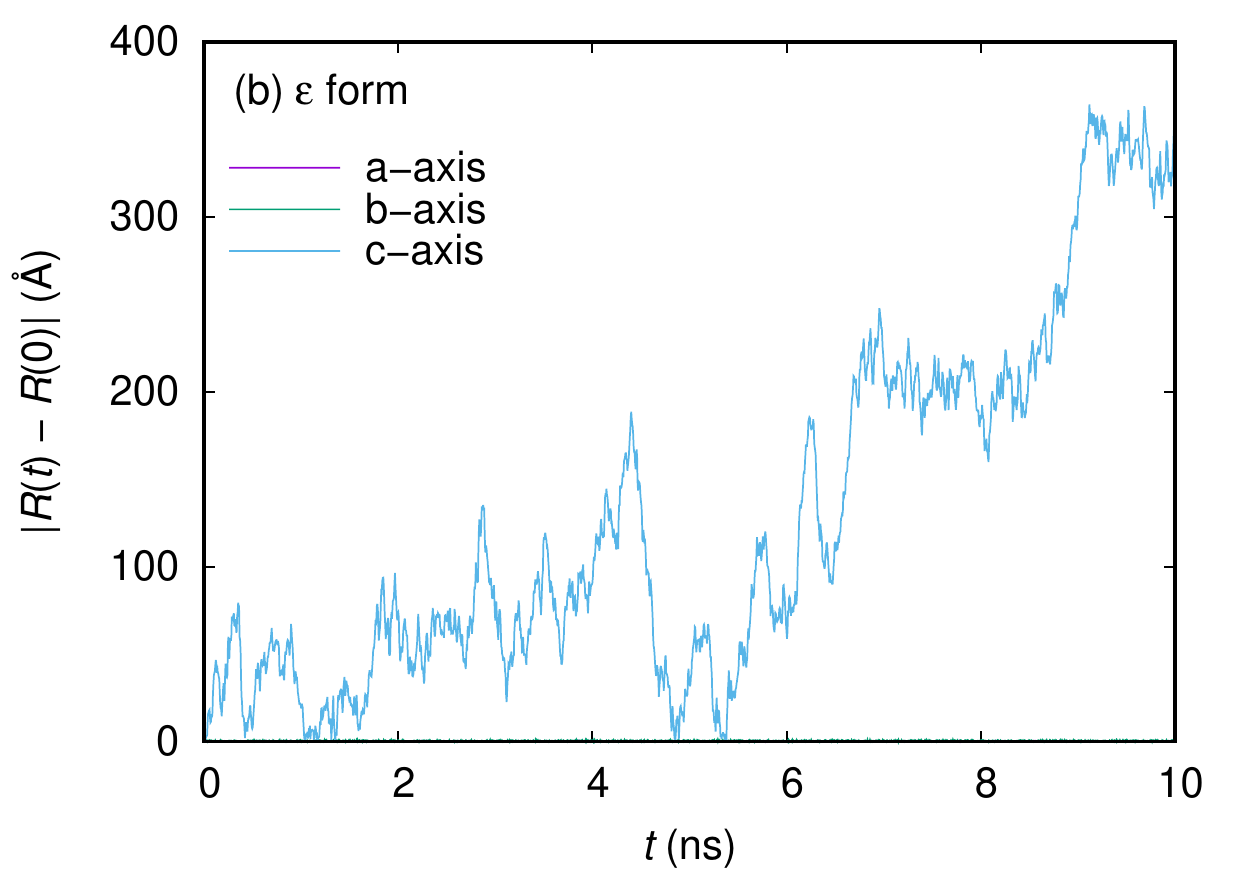}

  \caption{Displacement of one of CO$_2$ molecules along each axis
    in s-PS S-I (a) and $\varepsilon$ (b) form crystal.}
  \label{fig:dsp}
\end{figure}

\begin{figure}[ht]
  \centering
  \includegraphics[width=82mm]{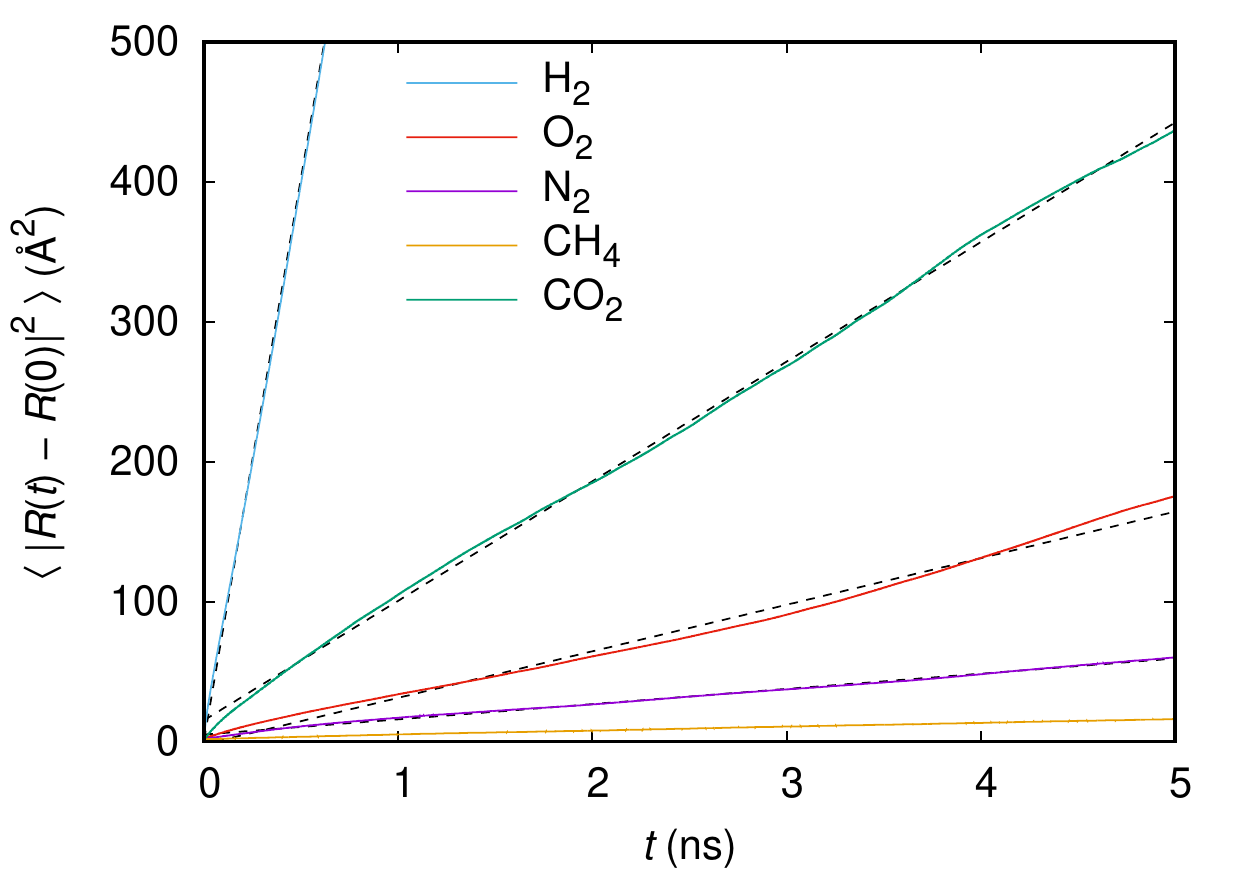}

  \caption{Mean-square displacement (MSD) of gases in s-PS S-I form crystal.
    The MSD values are fitted to linear lines (dashed lines)
    by the least-squares method.}
  \label{fig:MSD}
\end{figure}

The displacement of CO$_2$ in the s-PS crystal is shown in Fig. \ref{fig:dsp}.
The diffusion occurs only along the $c$-axis.
Its behavior is effectively one-dimensional.
As for the S-I form, 
the diffusion process consists of oscillation in the cavity 
and jump motion between neighboring cavities.
On the contrary, the jump motion is not so clear in the $\varepsilon$ form;
the gas transport is rather continuous, reflecting the wide channels.
The mean-square displacements (MSD) of the gas molecules 
are shown in Fig. \ref{fig:MSD}.
The H$_2$ molecule diffuses most quickly because it is the smallest molecule, 
while methane diffuses most slowly.

\begin{table}[ht]
  \caption{Diffusion coefficient of gases along $c$-axis, $D_c$, 
    in the S-I and $\varepsilon$ forms.
    Relative diffusion coefficient, $D_c / D_c^{\mathrm{CO_2}}$, 
    is also tabulated.}
  \label{tbl:D}

  \begin{center}
  \begin{tabular}{cccccc}
    \hline
    & \multicolumn{2}{c}{$D_c\times 10^{5}$ (cm$^2$/s)\textsuperscript{\emph{a}}}
    &
    & \multicolumn{2}{c}{$D_c / D_c^{\mathrm{CO_2}}$}
    \\
    \cline{2-3}\cline{5-6}
    \raisebox{1ex}[0pt]{gas} 
    & S-I form
    & $\varepsilon$ form 
    &
    & S-I form
    & $\varepsilon$ form
    \\
    \hline
    H$_2$  &   4.03~ (0.82)~ &  270 (74) &&  7.9~ & ~~6.9 \\
    O$_2$  &   0.144 (0.045) &  ~39 (13) &&  0.28 & ~~1.0 \\
    N$_2$  &   0.060 (0.015) &   48 (6)  &&  0.12 & ~~1.2 \\
    CH$_4$ &   0.008 (0.003) &  ~69 (13) &&  0.02 & ~~1.8 \\
    CO$_2$ &   0.51~ (0.22)~ &  ~39 (11) &&  1~~~ &   1   \\
    \hline
  \end{tabular}
  \end{center}

  \raggedright
  \textsuperscript{\emph{a}}Standard errors in parenthesis.\\
\end{table}

Diffusion coefficients were calculated from the slopes of the MSD.
Obtained diffusion coefficients are listed in Table \ref{tbl:D}.
Diffusion coefficients in the $\varepsilon$ form are approximately 
2 orders larger than those in the S-I form.
This is due to the wide tube-like channels in the $\varepsilon$ form.
In the $\varepsilon$ form, however, the 
difference in diffusivity between the gas species is very small, 
as indicated by the relative diffusion coefficients.
All the gas molecules can diffuse rapidly in the $\varepsilon$ form.
On the other hand, in the S-I form, 
the $D$ values are greatly different for different gas species.
The diffusion coefficient of CO$_2$ is ten times higher than that of N$_2$.
It is difficult to achieve such a sharp difference in diffusivity 
using amorphous membranes, 
because of their broad distribution of free volumes.

The remarkable point is that the tri-atomic molecule, CO$_2$, 
diffuses more quickly than the di-atomic molecules O$_2$ and N$_2$, 
in the S-I form.
In other words, larger molecules can diffuse more quickly in the S-I form.
This anomaly is not observed for the $\varepsilon$ form.

\subsection{Prediction of permeability and selectivity}

The permeability was calculated by
\begin{equation}
  P = DS, 
\end{equation}
on the basis of solution--diffusion model \cite{Wijmans1995}.
Table \ref{tbl:P-alpha} lists 
the predicted permeability $P$ and CO$_2$ selectivity.
Obviously, the solubility of CO$_2$ in the S-I form is higher 
than that of other gases.
Combined with effective diffusion of CO$_2$, 
the permeability of CO$_2$ in the S-I form becomes very high; 
approximately two hundred times higher than that of N$_2$.
This leads to higher CO$_2$ selectivity in the S-I form.
The CO$_2$ selectivity is very high in the S-I form, 
especially for the target gases N$_2$ and CH$_4$.
It is expected that the S-I membrane could be used 
for high efficiency carbon-capture systems 
installed in power plants (CO$_2$/N$_2$) or natural gas fields (CO$_2$/CH$_4$).

\begin{table}[ht]
  \caption{Predicted permeability $P$ and CO$_2$ selectivity 
    in the S-I and $\varepsilon$ forms.}
  \label{tbl:P-alpha}

  \begin{center}
  \begin{tabular}{ccccll}
    \hline
    & \multicolumn{2}{c}{$P$ (Barrer)\textsuperscript{\emph{a}}}
    &
    & \multicolumn{2}{c}{CO$_2$ selectivity\textsuperscript{\emph{b}}}
    \\
    \cline{2-3}\cline{5-6}
    \raisebox{1ex}[0pt]{gas} 
    & \multicolumn{1}{c}{S-I form}
    & \multicolumn{1}{c}{$\varepsilon$ form}
    &
    & \multicolumn{1}{c}{S-I form}
    & \multicolumn{1}{c}{$\varepsilon$ form}
    \\
    \hline
    H$_2$  & 1.4$\times 10^{3}$  &  9.2$\times 10^{4}$ &&  ~~5.6 &   10.2  \\
    O$_2$  & 3.8$\times 10^{2}$  &  1.4$\times 10^{5}$ &&  ~21   &  ~~7.0  \\
    N$_2$  & 4.5$\times 10^{1}$  &  1.5$\times 10^{5}$ &&  177   &  ~~6.4  \\
    CH$_4$ & 1.7$\times 10^{1}$  &  7.9$\times 10^{5}$ &&  460   &  ~~1.2  \\
    CO$_2$ & 8.0$\times 10^{3}$  &  9.5$\times 10^{5}$ \\
    \hline
  \end{tabular}
  \end{center}

  \raggedright
  \textsuperscript{\emph{a}}Barrer = 10$^{-10}$ cm$^3$(STP)cm/s cm$^2$cmHg.\\
  \textsuperscript{\emph{b}}Calculated by $P_{\mathrm{CO_2}}/P_{\mathrm{gas}}$.
\end{table}

\subsection{Comparison to experiments}

The permeabilities of gases in amorphous and semicrystalline $\alpha$ form
of s-PS were reported by Hodge et al. \cite{Hodge2001},
as tabulated in Table \ref{tbl:SDP-exp}.
Because amorphous s-PS is glassy state at room temperature, 
which is lower than glass transition temperature $T_{\mathrm{g}} \sim 370$ K, 
the diffusion coefficients are small in amorphous s-PS.
The $D$ values were reported to be enhanced 2--3 times
in the semicrystalline sample with crystallinity 0.3 \cite{Hodge2001}.
This enhancement was attributed to gas permeable channels 
in the crystals dispersed in amorphous regions.
The enhancement of permeability in semicrystalline samples 
compared to amorphous ones is also reported for 
poly(2,6-dimethyl-1,4-phenylene)oxide (PPO) \cite{Galizia2012},
which has also nanoporous-crystalline phase \cite{Tarallo2012,Rizzo2013}.
These behaviors are contrasted with gas diffusion process
in ordinary semicrystalline polymers such as polyethylene, in which
diffusion paths are limited to rubbery amorphous region \cite{Gusev1994}.

The calculated diffusion coefficients of CO$_2$ in s-PS S-I and $\varepsilon$ 
forms are 5.1$\times$10$^{-6}$ and 3.9$\times$10$^{-4}$, respectively, 
which are significantly large compared to that in amorphous s-PS,
4.0$\times$10$^{-8}$.
This is owing to the porous nature of these crystals, 
as evidenced by Fig. \ref{fig:fv-cluster} and \ref{fig:fv-PR}.

\begin{table}[ht]
  \caption{Experimental values of $S$, $D$, and $P$ of gases in s-PS 
    at 298 K \cite{Hodge2001}.}
  \label{tbl:SDP-exp}

  \begin{center}
  \begin{tabular}{cccc}
    \hline
      gas
    & $S$ (cm$^3$(STP)/cm$^3$atm)
    & $D$ (cm$^2$/s)
    & $P$ (Barrer)
    \\
    \hline
    \multicolumn{4}{c}{amorphous}\\
    O$_2$  & 0.17  &  1.1$\times$10$^{-7}$  & ~~2.5 \\
    CO$_2$ & 2.9~~ &  4.0$\times$10$^{-8}$  &  15  \\
    \multicolumn{4}{c}{$\alpha$ form crystal\textsuperscript{\emph{a}}}\\
    O$_2$  & 0.10  &  3.6$\times$10$^{-7}$  & ~~4.6 \\
    CO$_2$ & 1.8~~ &  8.2$\times$10$^{-8}$  &  17  \\
    \hline
  \end{tabular}
  \end{center}

  \raggedright
  \textsuperscript{\emph{a}}Crystallinity determined by X-ray diffraction is
  approximately 0.3.
\end{table}

\begin{figure}[ht]
  \centering
  \includegraphics{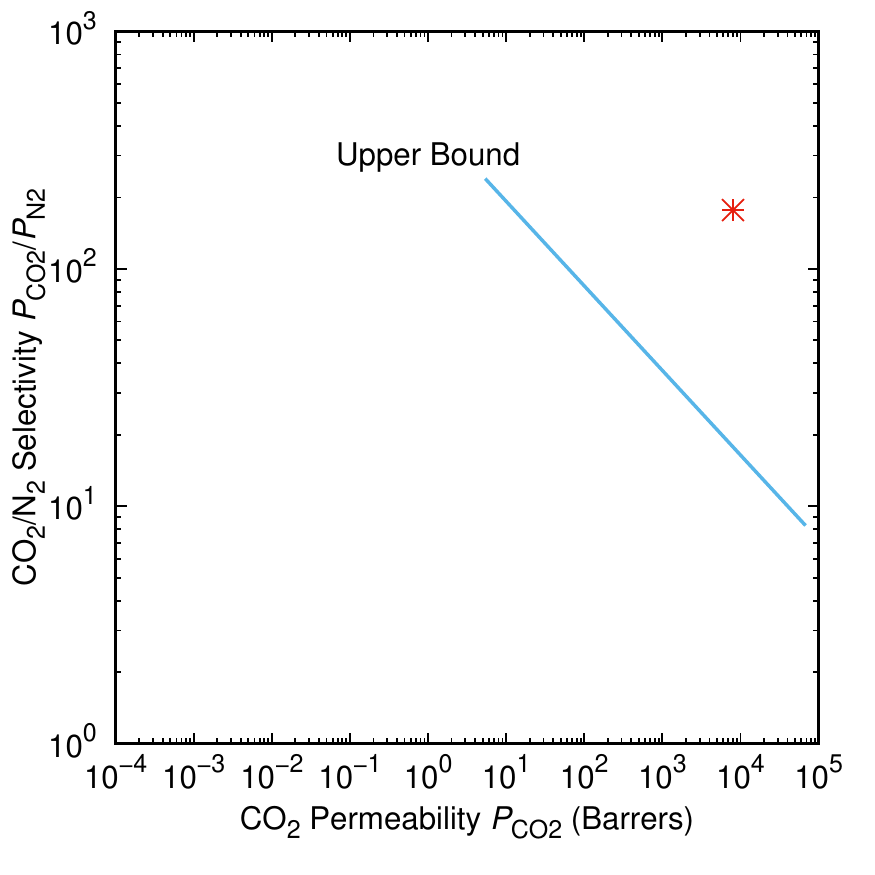}

  \caption{Upper bound correlation for CO$_2$/N$_2$ separation, 
    summarized by Robeson \cite{Robeson2008}.
    The CO$_2$ selectivity is plotted against permeability of CO$_2$
    for various commercial membranes.
    The star in the upper right indicates the predicted values 
    for the s-PS S-I form (this work).
  }
  \label{fig:Robeson-plot}
\end{figure}

The separation performance of the S-I form is compared to 
known commercial membranes in Fig. \ref{fig:Robeson-plot}.
Because of competing nature of permeability and selectivity, 
the commercial membranes are distributed below the upper-bound line
which has a negative slope.
The performance of the S-I form is far beyond the upper-bound 
as shown by the star in the figure.

\subsection{Origin of effective and efficient transport}

The solubility is mainly dependent on the interaction 
between the gas molecule and the host matrix.
Especially in the molecular cavities in the crystal, 
the repulsive interaction is severely affected by matching in 
the size and shape between guests and cavities.
Consulting Figs. \ref{fig:gas-model} and \ref{fig:fv-cluster}, 
the matching is good for CO$_2$, 
which is just fitted to the ellipsoidal cavity in the S-I form.
Because of the strong attractive interaction and 
the comfortable matching to the cavity shape and size, 
the solubility of CO$_2$ becomes high in the S-I form.

The diffusion process of small penetrants in dense amorphous polymers
have been understood as ``jump diffusion'', i.e., a series of hopping events
between voids \cite{Gusev1994}.
This kind of motion has been clearly observed by MD simulations.
The transition state theory has also been 
applied \cite{Gusev1994,GrayWeale1997,Greenfield2004}, in which 
the rates of solute's hops are dependent on the energy barrier 
(at the transition state) between voids.

\begin{figure}[ht]
  \centering
  \includegraphics[width=82mm]{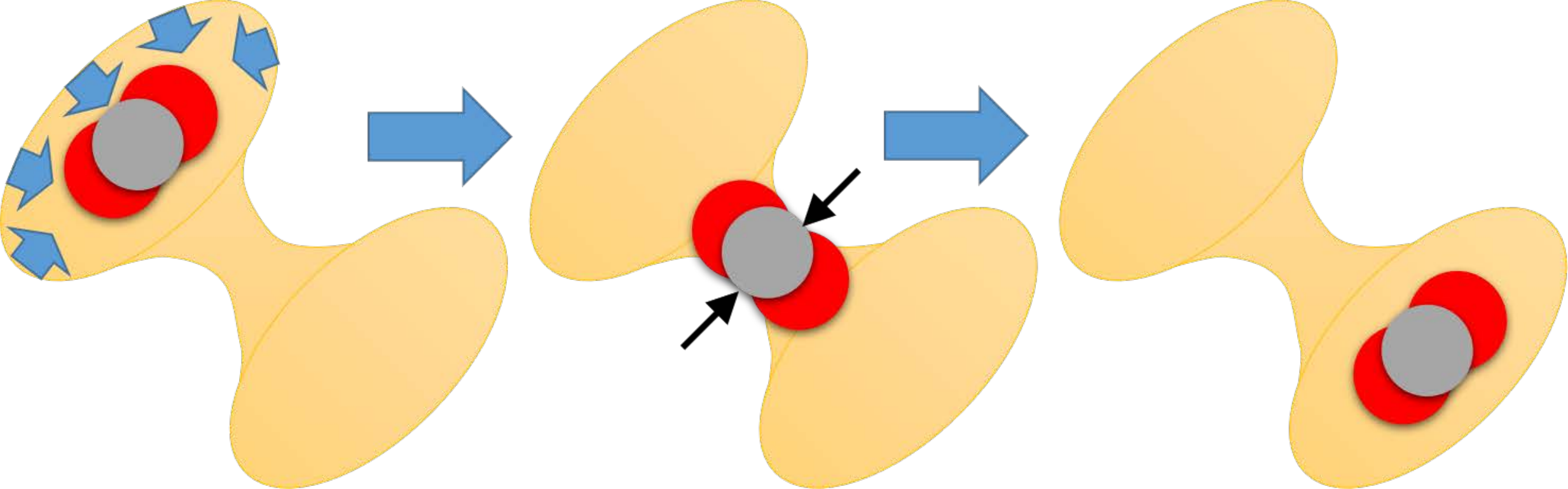}

  \caption{Jump-diffusion scheme of CO$_2$ in s-PS S-I form.}
  \label{fig:jump}
\end{figure}

In the s-PS S-I crystal, the diffusion scheme of gases is also jump diffusion,
as evidenced from the movies (Movie S1--S4) and 
the displacements of gases (Fig. \ref{fig:dsp}).
In contrast, the jump motion is not so clear in the $\varepsilon$ form,
where the gas transport is rather continuous.
The difference can be attributed to existence of energy barrier.
A hopping process in the S-I form is shown schematically in Fig. \ref{fig:jump}.
The CO$_2$ molecule is oscillating most of the time in the individual cavity,
interacting with host matrix (left and right panels), and
the jump motion sometimes occurs through the narrow channel
between neighbor cavities (middle panel).

In the S-I form, 
the diffusion process can be treated as a one-dimensional random walk.
The jump length can approximately be treated as constant, 
defined by crystalline periodicity.
Therefore, 
the diffusion coefficient is dependent solely on the jump probability.
The jump probability is mainly governed by 
the transition state
(middle panel of Fig. \ref{fig:jump}) of individual jump event, and 
expected to be proportional to the Boltzmann factor of the activation energy.
The activation energy of the jump event is thought to be affected 
by the short-axis length of the gas molecule.
Although the long-axis length of CO$_2$ is the largest, 
the short-axis length is smaller than that of N$_2$ or CH$_4$, 
and almost the same as that of O$_2$.
Therefore, the diffusion coefficient of CO$_2$ becomes larger 
than that of N$_2$ or CH$_4$.

\begin{figure}[ht]
  \centering
  \includegraphics[width=82mm]{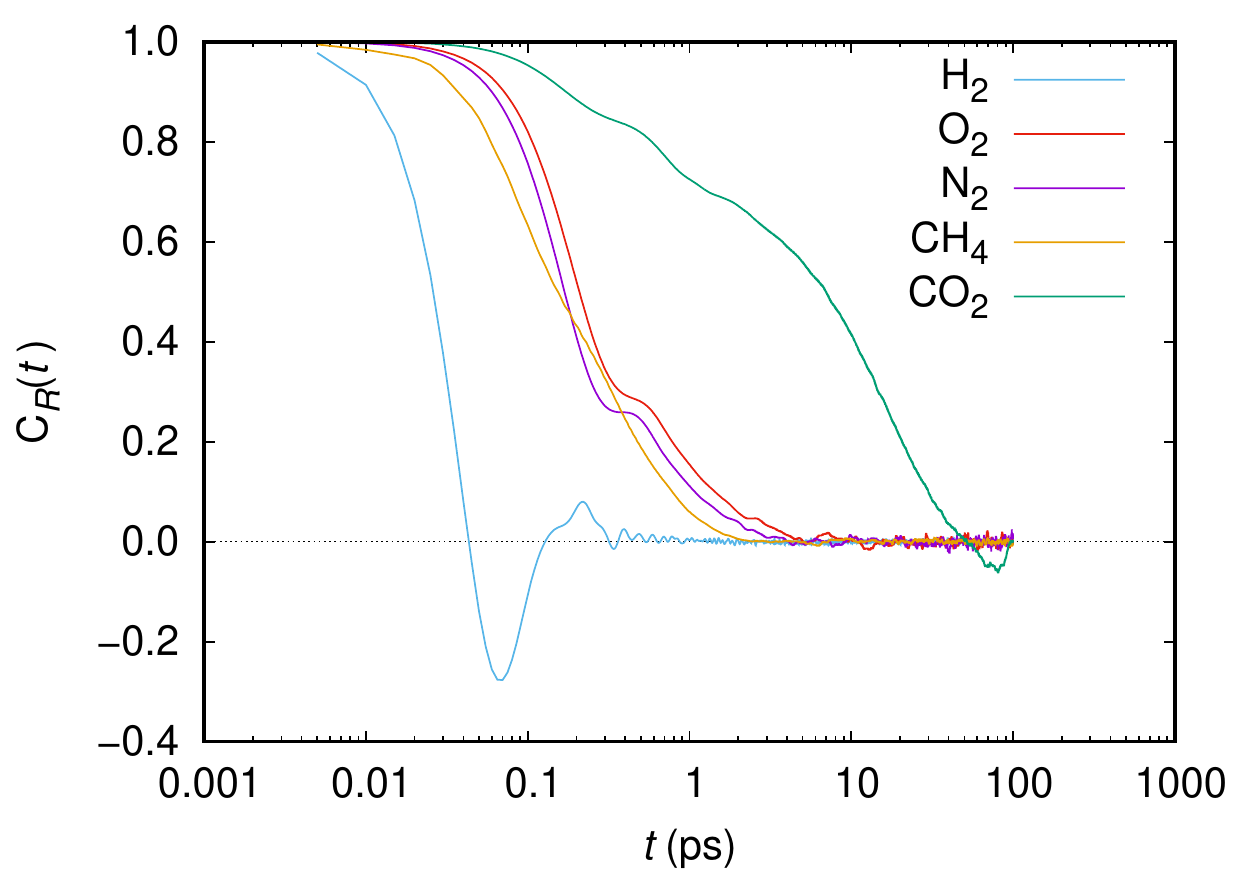}

  \caption{Re-orientational auto-correlation function of gases 
    in s-PS S-I form crystal.}
  \label{fig:RAF}
\end{figure}

However, the latter raises an important question.
Why is it that the diffusion coefficient of CO$_2$ is four times larger 
than that of O$_2$, in spite of their similar diameters?
Please recall the animation of diffusion behavior of CO$_2$ and N$_2$
in Supporting Information.
CO$_2$ molecules are oscillating in the cavity, 
while N$_2$ molecules are rotating.
This is because the full length (long-axis length) of the CO$_2$ molecule 
is too long to re-orient in the ellipsoidal cavity.
From the re-orientational auto-correlation function of gases, 
Fig. \ref{fig:RAF}, 
the re-orientational relaxation time of CO$_2$ is confirmed to be
two orders longer than those of O$_2$ and N$_2$.

From these observations, 
the author proposes that the momentum transfer from polymer matrices is effectively 
delivered to the translational degree of freedom for a tri-atomic molecule.
For di-atomic molecules, 
the impulses from polymer chains are transferred into angular momentum.
The gas molecules spin in the cavity.
Therefore, the impulses from the matrices do not lead to jump trials.
In contrast, for tri-atomic molecules, 
the transferred momentum is effectively used for translation,
because CO$_2$ molecules cannot rotate in the cavity.
In other words, 
CO$_2$ molecules more frequently exchange their momentum to the host matrix.
This leads to an increase in jump trials.

\subsection{Proof by ``long-gas'' model}

\begin{figure}[ht]
  \centering
  \includegraphics[width=82mm]{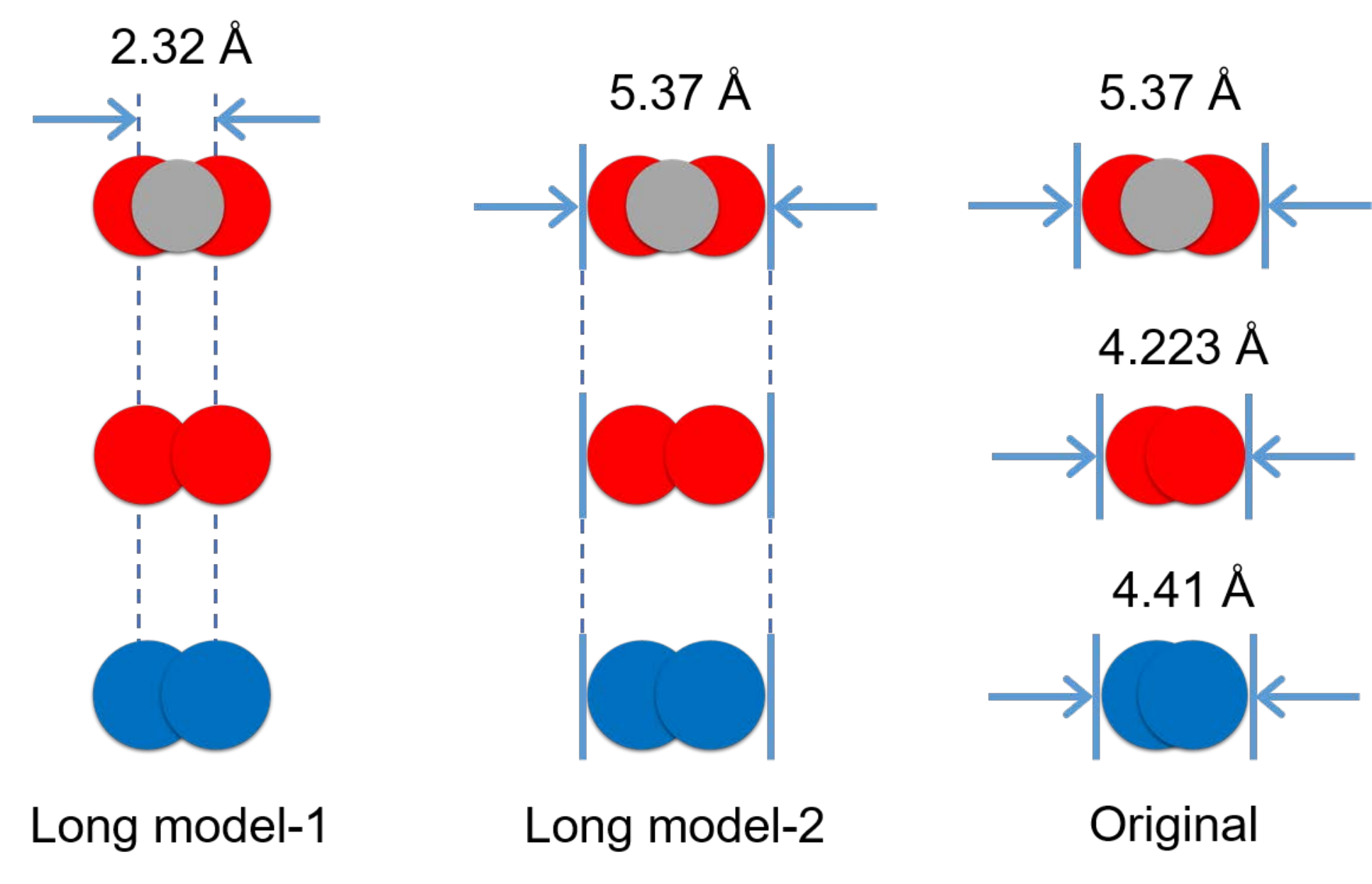}

  \caption{``Long-gas'' model of O$_2$ (middle) and N$_2$ (bottom),
    compared with CO$_2$ (top).}
  \label{fig:long-gas}
\end{figure}

\begin{table}[ht]
  \caption{
    Force-field parameters for ``long-gas'' model
  }
  \label{tbl:FF-parm-long}

  \begin{center}
  \begin{tabular}{ccllcll}
    \hline
    & 
    & \multicolumn{2}{c}{model-1\textsuperscript{\emph{a}}}
    &
    & \multicolumn{2}{c}{model-2\textsuperscript{\emph{b}}}
    \\
    \cline{3-4}\cline{6-7}
    \raisebox{1ex}[0pt]{gas} 
    & \raisebox{1ex}[0pt]{atom}
    & \multicolumn{1}{c}{$q$ (e)}
    & \multicolumn{1}{c}{$l_{\textrm{b}}$ ({\AA})}
    &
    & \multicolumn{1}{c}{$q$ (e)}
    & \multicolumn{1}{c}{$l_{\textrm{b}}$ ({\AA})}
    \\
    \hline
    N$_2$  & N & $-$0.1084 & 1.16 && $-$0.1374 & 1.03  \\
           & M & $+$0.2168 &      && $+$0.2748 & \\
    O$_2$  & O & $-$0.0350 & 1.16 && $-$0.0324 & 1.1785\\
           & M & $+$0.0670 &      && $+$0.0648 & \\
    \hline
  \end{tabular}
  \end{center}

  \raggedright
  \textsuperscript{\emph{a}}Same bond length as CO$_2$.\\
  \textsuperscript{\emph{b}}Same full-length as CO$_2$.
\end{table}

In order to substantiate this hypothesis, an
MD simulation was performed using the ``long-gas'' models 
shown in Fig. \ref{fig:long-gas}.
In model-1, O--O and N--N bond lengths 
(originally 1.21 and 1.10 {\AA}, respectively) are elongated to 
the same length as C--C of CO$_2$ ($2l_{\textrm{b}}$ = 2.32 {\AA}).
In model-2, bond lengths of di-atomic molecules are adjusted so that 
their molecular full lengths are identical to that of 
CO$_2$ ($2l_{\textrm{b}} + \sigma_{\textrm{OO}}$ = 5.37 {\AA}).
The partial charges of atoms were adjusted to reproduce quadrupole moments
of original models.
The model parameters are listed in Table \ref{tbl:FF-parm-long}.

The calculated diffusion coefficients for the long-gas model 
are listed in Table \ref{tbl:D-long-gas}.
Using the model, the diffusion coefficients were surely enhanced, 
approximately 5--10 times.
The $D$ values of O$_2$ and N$_2$ calculated for the long model 
were approximately the same as that of CO$_2$, 
0.51$\times 10^{-5}$ cm$^2$/s.
It is revealed that matching of molecular structure with cavity shape is
important for control of diffusivity in the crystalline membrane.

\begin{table}[ht]
  \caption{Diffusion coefficients calculated for ``long-gas'' models.}
  \label{tbl:D-long-gas}

  \begin{center}
  \begin{tabular}{cccc}
    \hline
    & \multicolumn{3}{c}{$D_c\times 10^{5}$ (cm$^2$/s)} \\
    \cline{2-4}
    \raisebox{1ex}[0pt]{gas} & model-1 & model-2 & original \\
    \hline
    \multicolumn{4}{c}{S-I form}\\
    O$_2$  & 0.61 (0.07) & 0.70 (0.07) & 0.144 \\
    N$_2$  & 0.61 (0.15) & 0.32 (0.05) & 0.060 \\
    \multicolumn{4}{c}{$\varepsilon$ form}\\
    O$_2$  &  43 ~(2)    &  43 (5)     &  39 \\
    N$_2$  &  55 (13)    &  43 (4)     &  48 \\
    \hline
  \end{tabular}
  \end{center}
\end{table}

\subsection{Design criteria for crystalline membrane}

These findings suggest a novel design criteria 
for CO$_2$ separation membranes using porous polymer crystals.
For crystalline membranes, 
it is important to control mesh size and affinity, 
which affect both diffusivity and solubility.
In addition, diffusivity can be controlled by matching in shape 
between cavity and gas molecule.
Indeed, in some cases, larger molecules can diffuse more quickly;
diffusion coefficient of CO$_2$ is larger than that of O$_2$ and N$_2$.
As evidenced from the present simulation, 
the momentum of the polymer matrix is effectively transferred 
to the translational degree of freedom of CO$_2$ molecules.
A novel aspects, 
controlling the momentum transfer between matrices and penetrants
by coupling to the re-orientational motion of gases in regular cavities,
could be introduced into the design criteria.

For amorphous membranes, as contrasted with the crystalline membranes, 
separation by diffusivity is insufficient 
because of the broad distribution of 
the activation energy at the transition state.
Furthermore, it is difficult to control the momentum transfer 
of gas molecules in amorphous polymers
because of a wide variety of cavity size and shape.

From a technological point of view, 
we have to overcome the difficulty that 
the crystallinity is generally low (up to 50 \% or so) 
and complex spherulites are formed for polymer systems.
The best performance of separation, presented in this article, 
can be achieved for oriented single crystals.
The solvent crystallization on the solid surface may be promising
to obtain oriented single crystals; 
the crystal size and orientation of the s-PS $\delta$ form
can be controlled by solvent species, 
using host--guest interactions between polymer and solvents \cite{Daniel2006}.
If once the oriented single crystal of the $\delta$ form is obtained, 
it may be transformed into the $\varepsilon$ form by a solvent treatment 
and further transformed into the S-I form by applying stress.
A confined crystallization in polymer nanolayered films 
may also be used to obtain oriented nano-crystals \cite{Carr2012}.

Note that the design criteria proposed above 
is not specific to the s-PS S-I form, but may be widely applicable
to the design of separation membranes with ordered structures.
One of the limitation of the S-I form to use for gas separation
is that it needs stress to keep its structure.
It is challenging to search for other crystals with similar cavity structures 
even under atmospheric pressure, by applying the design criteria.
Following the design criteria, 
the permeability of other tri-atomic molecules, 
for example, N$_2$O (full length $\sim$5.4 {\AA}, known as greenhouse gas), 
may be enhanced; N$_2$O may be effectively captured from the air.

\section{Conclusions}

The crystal structure of a new orthorhombic ``S-I'' form of s-PS 
under uniaxial stress ($\sigma_{yy}$ = 0.27 GPa) 
was determined by MD simulations.
The space group is $Pbcb$ and the lattice constants are 
$a$ = 10.96 {\AA}, $b$ = 16.55 {\AA}, and $c$ = 7.87 {\AA}.
The density is 0.969 g/cm$^3$, which is almost the same as that of
the empty $\delta_{\mathrm{e}}$ form.
In the S-I form, chains with relaxed L and L (or R and R) contacts 
form zigzag channels at the molecular level.

It was predicted that the S-I membrane could have
significantly higher CO$_2$/N$_2$ ($\sim$180) and 
CO$_2$/CH$_4$ ($\sim$500) separation factors 
with preserving its high CO$_2$ permeability 8000 Barrers.
The mechanism of the effective CO$_2$ transport in the S-I form 
was elucidated in detail.
The CO$_2$ molecule is just fitted to the ellipsoidal cavities in the S-I form
and effectively interacts with surrounding polymer matrix. 
This leads to the higher solubility of CO$_2$.

In addition to this, 
tri-atomic molecules can more effectively diffuse in the S-I form, 
compared with di-atomic molecules.
The CO$_2$ molecule is oscillating, 
while the N$_2$ and O$_2$ molecules are rotating in the ellipsoidal cavities.
The momentum from the polymer matrix is effectively transformed into
translational motion for CO$_2$, 
while it is transferred into rotational degree of freedom for N$_2$ and O$_2$.
As a result, the diffusion coefficient of CO$_2$ becomes higher
than that of N$_2$ and O$_2$.
The effect of long-axis length of gas molecules on diffusivity 
was confirmed by the ``long-gas'' model, 
in which the lengths of N$_2$ and O$_2$ were elongated 
to the same length as that of CO$_2$.
The diffusion coefficients of di-atomic molecules were surely enhanced
by the long-gas model.

Because of the enhanced solubility and diffusivity, 
the permeability of CO$_2$ becomes extremely higher in the S-I form; 
CO$_2$ can be effectively separated from the mixed gas with N$_2$ or CH$_4$.
To design high-efficiency CO$_2$ transport membranes, 
it is important to control the momentum transfer from polymer matrix
to the translational degree of freedom of the penetrant.

\section*{Acknowledgments}
This work was partly supported by JSPS KAKENHI, grant number 19550121.
The calculations were performed on supercomputer at ACCMS, Kyoto University.
The fee of the supercomputer was supported by CII, University of Fukui.

\appendix
\section{Calculation of X-ray intensity}
\label{sec:intensity}
The structure factor $F_{hkl}$ for the Miller index $(hkl)$ was calculated by
\begin{equation}
  F_{hkl} 
  = \sum_{j} f_j T_j \exp\left[ 2 \pi i (h x_j + k y_j + l z_j) \right],
\end{equation}
where $f_j$ is the atomic scattering factor of atom $j$, 
$T_j$ is the temperature factor ($B$ = 8 {\AA}$^2$), 
$i$ is the imaginary unit, and 
$(x_j, y_j, z_j)$ is the fractional coordinate of atom $j$.
Hydrogen atoms were also included in the calculation.
The reciprocal lattice vector 
$\mathbf{k} = h\mathbf{a}^* + k\mathbf{b}^* + l\mathbf{c}^*$, 
where $\mathbf{a}^*$, $\mathbf{b}^*$, and $\mathbf{c}^*$ are 
the reciprocal primitive vectors, 
is related to the diffraction angle $2\theta$ by:
\begin{equation}
  |\mathbf{k}| = \frac{ 2 \sin\theta }{ \lambda }, 
\end{equation}
where $\lambda$ = 1.5418 {\AA} is the wavelength of the Cu K$\alpha$ radiation.
The $f_j$ value is approximated by:
\begin{equation}
  f
  = \sum_{p=1}^{4} a_p \exp\left( - \frac{b_p \sin^2\theta}{\lambda^2} \right)
  + c,
\end{equation}
where the parameters $a_p$, $b_p$, and $c$ are published for each element
\cite{ITC-Vol.4}.
The $F_{hkl}$ values were calculated for sets of $h$, $k$, and $l$ 
satisfying $\sqrt{h^2 + k^2 + l^2} \leq 7$.
The X-ray diffraction intensity $I_{hkl}$ was calculated by:
\begin{equation}
  I_{hkl} = |F_{hkl}|^2 Lp,
\end{equation}
where $Lp$ is the Lorentz polarization factor for X-ray powder diffraction:
\begin{equation}
  Lp = \frac{1 + \cos^2{2\theta} }{ 2 \sin^2{\theta} \cos{\theta} }.
\end{equation}

\section{Supplementary data}
Supplementary material related to this article can be found online
at https://doi.org/10.1016/j.memsci.2021.120202.


\end{document}